\documentclass{article}
\pdfoutput=1 %%%%%%%%%%%%%%%%%%%%%%%%%%%
\usepackage{latexsym}
\usepackage{amsmath,amssymb}
\usepackage{float}
\usepackage[dvipdfmx,hiresbb]{graphicx}
\begin{document}

\pagestyle{plain} 
\setcounter{page}{1}
\setlength{\textheight}{650pt}
\setlength{\topmargin}{-40pt}
\setlength{\headheight}{0pt}
\setlength{\marginparwidth}{-10pt}
\setlength{\textwidth}{20cm}

\title{Promotion of Cooperation in Coevolutionary Public Goods Game on Complex Networks with and without Topology Change}
\author{Norihito Toyota\\ %\and 
Faculty of Business Administration and Information Science, \\ \hspace{6mm} Hokkaido Information University, Ebetsu, Nisinopporo 59-2, Japan\\
toyota@do-johodai.ac.jp }
\date{}
\maketitle
%%%%%%%%%%%%%%%%%%%%%%%%%%%%%%%%%%%%%%%%%%%%%%%%%%%%%%%%%%%%%%%%%%%%%%%%%%%%%%%%%%%%%%%%%%%%%%%%%
%
% LaTeX2e'ÆLaTeX209"»•Ê'Ì'½'ß'Ì'è‹`iˆÈ‰º3sj'͉Ȍ¤"ïƒ}ƒNƒkkhgrp.mac
% 'Ì'è‹`'ðCìŽÒ'Ì‹à'ò'åŠwÂ–ؐ搶'Ì‹–‰Â'𓾂ė˜—p'³'¹'Ä'¢'½'¾'«'Ü'µ'½D
%
%&&&\newif\ifLaTeXe\LaTeXefalse
%&&&\expandafter\ifx\csname PackageError\endcsname\relax\LaTeXefalse
%&&&\else\LaTeXetrue\fi

%&&&\ifLaTeXe
%&&&  \documentclass{article}
 %%%  \usepackage{SICE-SSI}
 % \usepackage[dvips]{graphicx}
%&&&  \usepackage[dvipdfm]{graphicx}%%%%%%%%%%%%%%%%%%%%%%%%%%%%%%%%%%%%%%%%%%%%
%&&&\else
%%  \documentstyle[SICE-SSI,epsbox]{jarticle}
%&&&\fi

%&&&\begin{document}
%&&&\title{A study of Inverse Ultra-discretization of  cellular automata }
%&&&\author{Norihito Toyota\ Hokkaido Information University }

\abstract{
 The evolution of cooperation among unrelated individuals in human and animal societies remains a challenging issue across disciplines. 
It is an important  subject also in the evolutionary game theory to understand how cooperation arises. 
The subject has been extensively studied  especially in Prisoners' dilemma game(PD) but 
the emergence of cooperation is also an important subject in public goods game(PGG).  
 In this article, we consider coevolutionary PGG on complex networks where both the topology of the networks and strategies that players adopt under the influence of game dynamics varies.  
Though cooperators can contribute a fixed amount per game in PGG on networks in the previous studies,  the cooperators contribute a fixed amount per member of the group in PGG of this article. 
The latter is seemed  to be more natural than the former model. 
 These models lead to great differences in both PGGs. 
We study what effects on the evolution of player's strategies, defection and cooperation and the average payoff does the interaction between the game dynamics and the network topology bring. 
Moreover by comparing the models that do not depend on game dynamics to the models without topology changing, we intend to uncover the effect of the topology chnage and the game dynamics on the promotion of the cooperation.  
As result we intend to clear in what situations cooperation strategy is promoted or preserved by investigating them by making computer simulations. 
We also investigate the relation between the ratio of the cooperator and  the average payoff over all players. 
Furthermore we study how  do the topology of initial  networks have an influence on the cooperation and the average payoff. 

\subsubsection*{keyword; public goods game, complex networks, evolutionary game, cooperation,  \\ \hspace{17mm}network evolution }

%&&&\maketitle\thispagestyle{empty}
%&&&\pagestyle{empty}
%%%%%%%%%%%%%%%%%%%%%%%%%%%%%%%%%
\normalsize
%%%%%%%%%%%%%%%%%%%%%%%%%%%%%%%%%%

\section{Introduction}
The evolution of cooperation among unrelated individuals in human and animal societies remains a challenging issue across disciplines. 
Since the 1990s, it has been an important  subject also in the evolutionary game theory\cite{Smi} to investigate how cooperation arises. 
The subject has been extensively studied during the past thirty years, especially in Prisoners' dilemma game(PD) and its various modified versions \cite{Now}.
A rational player chooses the Nash equilibrium strategy, defection, in the original version of the one-off PD with two strategies by two players\cite{Wei}.  
So it is a long-standing  question by what mechanism cooperation, that is Pareto Optimum, is realized.   
In the iterated  PD, a tendency to choose the Nash equilibrium strategy is relieved under some conditions\cite{Ax}. 
There are many circumstantial evidences that the cooperation is promoted within groups in the evolutionary PD on lattice, which is a sort of spacial games\cite{Now},\cite{Now2}.

Corresponding studies for evolutionary games on complex networks\cite{New}, which are considered as metaphors for various relations in the real world, in addition to regular networks such as lattice, have been investigated\cite{Sza}, \cite{Dor},\cite{Santos1}. 
%\bibitem{Zim1}M.G.Zimmermann, V.M.Eguiluz and M.S. Miguel,"Cooperation, adaption and the emergence of leadership" In J.B.Zimmermann and  A.Kirman, Economics and Heterogeneous Interacting Agents, Springer, Berlin. pp.73-86, 2001
In \cite{Zim1}, first the effect of a network topology on the promotion of cooperation has been studied in games with two players. 
 Moreover it is begun to pay its attention to the relations between the evolution of cooperation and the network structure in social dilemma games\cite{Perc1}\cite{Holme}. 
The researches on how does game dynamics on networks influence the network topology, has been developed\cite{Fu},\cite{Gross}. 

 %\bibitem{Zim2}  M.G.Zimmermann, V.M.Eguiluz and M.S. Miguel,"Cooevolution of dynamical states and interactions in dynamic networks", Phys. Rev.E69,065120,2004.
\cite{Zim2},\cite{Pach1} studied the coevolution of iterated games and showed that the heterogeneity of networks\cite{Zim3} such in the scale free networks\cite{Bara}, the lifetime of links\cite{Pach2}, the ratio of the time scale of the structure  change to that of strategy change\cite{Santos11}  and so on are of importance for the promotion of cooperation.  
%\bibitem{Zim3} M.G.Zimmermann, V.M.Eguiluzl,"Cooperation, Social networks and the emergence of leadership in Prisoner's dilemma with local interactions", Phys. Rev.E72,056118, 2005, 
%\bibitem{Pach1}J.M. Pacheco, A.Traulsen and M.A.Nowark, "Active linking in evolutionary games", J.Theor.Boi.243,pp.437-443,2006.
%\bibitem{Pach2}J.M. Pacheco, A.Traulsen and M.A.Nowark, "Coevolution of strategy and structure in complex networks with dynamical linking", Phys.Rev.Lett.97,258103,2006. 
%\bibitem{Santos1}F.C.Santos, J.M.Pacheco and T.Leneaerts, "Cooperation previles when individuals adjust their social ties", PLoS. Compt. Biol.2, pp.1284-1290, 2006. 
The interplay between the network structures and game dynamic with respect to the promotion of the cooperation  has been discussed\cite{Fu10},\cite{Lee}. 
%\bibitem{Fu10}F.Fu, X.Chen, L>Liu, L.Wang, "Promotion of cooperation induced by the interplay between structure and game dynamics", Physica A383,pp.651-659,2007. 

In coevolutionary models, the models with topology change independent of strategy have been studied\cite{Perc10}, \cite{kun},\cite{Tani}. 
There, it was also shown that growing networks with a middle value of a maximum degree also leads to heterogeneous networks\cite{Rong1} and promotes  cooperation\cite{Perc2}. 
%\bibitem{Szol1}A. Szolonoki, M.Perc and Z. Panku,"Making new connections towards cooperation in the Prisoner's Dilemma game", EPL. 84, 50007,2009.
%\bibitem{Rong10}Z.Rong, X.Li, X.Wang, "Roles mixing patterns in cooperation on a scale free networked game", Phys. Rev. E79,027101, 2007
%\bibitem{Perc10}M.Perc, A.Szolnoki and G.Szabo, "Restricted connections among distinguished players support cooeration",  Phys.Rev. E78,066101,2008
%\bibitem{kun}A.Kun and I.Scheuring, "Evolution of cooperation on dynamical graphs", BioSysterm 96, pp.65-68,2009
%\bibitem{Tani} J.Tanimoto, "The effect of assortative mixing on emergency cooperation in an evolutionary network game", In:IEEE Conference on Evolutionary Computation, pp.487-493,2009
 Moreover the PD on growing networks\cite{Pon1}\cite{Pon2}, time-varying networks and other networks \cite{Car}  for promoting cooperation have been studied.  
 Recently it is pointed that the evolution of strategies only may not  satisfyingly promote cooperative behavior in  the evolutionary games. 
It has been rather recognized that the effect of  spatial structure and heterogeneity in addition to the evolution of strategies is important for cooperation \cite{Perc2},\cite{Perc3},\cite{Perc4}. 
In the coevolution in the evolutionary games, another property besides the evolution of strategies may simultaneously evolves in general.  
See an excellent review by \cite{Perc5} for researches about such coevolution. 
Furthermore it has been pointed out that the combination of network structure and other factors such as memory promotes cooperation behavior\cite{Hadz1}, \cite{Hadz2}. 
%%%%%%%%%%%%%%%%%%%%%%
However, its assertion are not yet conclusive and sufficiently general, because the analysis of the interaction between network topology and game dynamics is rather intricated\cite{Toyo}. 
%There, however, are not confident and general  assertions yet, because the analysis of interactions between network topology and network dynamics is rather intricated \cite{Toyo}.
%%%%%%%%%%%%%%%%%%%%%%%%%% 

On the other hand, the Public Goods Game(PGG) \cite{Now},\cite{Mc},\cite{Hau} that can be interpreted as $n$-persons PD has been considered.  
Studies for cooperation as have been made in PD game have already begun  also in PGG\cite{Bat}. 

There are some studies of PGG on fixed networks\cite{SPerc}. 
First it has been shown that the social diversity in networks play a crucial role to promote cooperation by some researchers\cite{Santos2}
,\cite{Yang10},\cite{Cao}.
%\bibitem{Yang10}H-X.Yang, W-X.Wang, Z-X Wu and B-H. Wang,  "Diversity optimized cooperation on complex networks", Phys.Rev. E79, 056107, 2009.
%\bibitem{Cao}X-B. Cao, W-B. Du and Z-H. Rong, "Evolutionary public goods game on scale-free networks with heterogeneous invest ment", Physica A389, pp.2273-1280, 2010.
The effect of assortativity in scale-free networks on the cooperation has been also explored in \cite{Rong10}.
%\bibitem{Rong10} Z. Rong and Z-X.Wu, "Effect of the degree correlation in public goods game on scale-free networks", EPL.87, 30001, 2009. 
Furthermore the authors  of \cite{Rong10} showed that the clustering of networks has a beneficial effect on the promotion of cooperation as it favors the formation and the stability of compact cooperative clusters. 
The impact of degree-correlated aspiration levels has also been studied in \cite{Yang11}, \cite{Zhang1}. 
%\bibitem{Yang10} H-X.Y. Yang, Z.Rong, P-M. Lu and Y-Z Zeng,  "Effect of aspiration on public cooperation in structured populations", Physica A.391, pp.4043-4049, 2012
%\bibitem{Zhang10} H-F. Zhang, R-R. Liu, Z.Wang, H-X. Yang and H-B. Wang, "Asoiration-induced reconnection in spacial public-goods game", EPL. 94, 18006, 2011
 %%AIMAI%

Especially PGG on complex networks with topology change has been studyed\cite{Zhang1},\cite{Wu1},\cite{Zhang2},\cite{Sha},\cite{Perc6}. 
 PGG using the normalized enhancement factor $\delta$  on random networks\cite{Ren} has been studied in  \cite{Wu1}.  
A focal player compares his/her payoff  with the payoff of a player chosen randomly among neighbors  and change the strategy based on Fermi rule where either the focal player experience strategy change with a probability $w$ or network change is undergone with a probability $1-w$. 
In \cite{Wu2}, PGG on Barabasi-Albert network\cite{Bara} has been studied. 
There the network is fixed but a focal player plays PGG with  $g$ players, which is a fixed number,  among neighbors of the focal player.   
In the model, comparing the focal player's payoff  with the payoff of a player chosen randomly among neighbors,  his/har strategy is changed  and, the change of the group structure is based on "reputation" of players.

In PGG on a lattice \cite{Zhang1},  players break the adverse social ties bringing the worst productivity, whereas remain ties with others. 
After breaking the ties, the player will search for a more beneficial interaction with another partner among the neighbors of neighbors.
There  a measure of individual's inertia to react to their rational selections both at strategy and topological levels is introduced. 
The strategy  of a node and the local network structure are updated with probabilities $\omega$ and $1-\omega$, respectively. 
Different $\omega$ values separate those so that the two time scales is implemented by asynchronous updating $\omega$. 
The model in \cite{Zhang2} has explored PGG on Newman-Watts small world network\cite{New2}. 
A characteristic property of their model is that  the change of network topology is undergone based on "aspiration level",  whereas the strategy is undergone based on Fermi rule. 
The model proposed by \cite{Sha} that introduced "aspiration" for strategy change and the change of network topology where an unsatisfied  player with negative satisfaction defined by using the aspiration breaks an edge to random neighbor, changes strategy or add an edge to a random player.   
The model is only based on the player's personal information but doesn't need to know any information of the neighbors of 
the player.   
All of the works assert that cooperation can be promoted or maintained within the framework of their models. 
A fine review including rich references is described in \cite{Perc7}, also \cite{Sza} and \cite{Perc9}.

  In this article, we consider an evolutionary PGG on diverse complex networks where the topology of the networks varies mainly  under the influence of game dynamics.  
There the strategies that players employ also change under the influence of game dynamics.  
The proposed models are so simple in the sense where any stochastic change of strategies and any outside factors such as the aspiration (level), Fermi rule and  reputation are not introduced and furthermore we try not to introduce any factors to promote explicitly C-strategy like a punishment to the utmost;.  
We consider some evolutionary strategy models that are thought to reflect and simplify people's features  in the real world. 
The models sometimes include a part of futures of above-mentioned models\cite{Wu1},\cite{Wu2},\cite{Zhang1},\cite{Zhang2}, \cite{Sha},\cite{Perc6}.  
Moreover we will compare and study these models in a lump. 
 The some models reflect game dynamics whereas others  do not in the change of strategies and the network structure. 
Though diverse complex networks are investigated as initial networks, we mainly study how does the interaction between the game dynamics and the network topology change influence the evolution of player's strategies, defection and cooperation, and the average payoff over all players.   
Investigating them by making computer simulations, we intend to clarify in what situations cooperation is promoted or preserved. 
We also intend to investigate the influence of the interaction on the average payoff over all players. 
The corresponding studies have been already investigated in \cite{Toyo2}.  
Though these cooperators can contribute a fixed amount per game in PGG on networks in \cite{Toyo2}, cooperators contibute a fixed amount per member of the group in PGG of this article. 
This is a great difference in  both PGGs. 
The latter is seemed  to be more natural than  the former model,  
because it is assumed that players implicitly have infinite assets in the former model 
where a player plays the degree $k$ of the player $+1$ PGGs. 
Thus as $k$ becomes large, the asserts that the player has must becomes large.   
This assumption is though to be quite unrealistic. 
In the meanwhile, the latter means implicitly that players have some finite asserts. 
     
%\\aaaaaaaaaaaaaaaaaaaaaaaaaaaaaaaaaaaaaa\\

The structure of this article is as follows. 
In the section 2, we explain the original PGG and the PGG on a network. 
After that we introduce PGG on complex network as our models.  
There we introduce evolutionary models with diverse tactics including network evolution. 
The some models  reflect game dynamics whereas others  do not. 
Then the comparison of  both type models would uncover the influences of game dynamics on coopperation or affluence. 
Even more some models change the network topology every game round where the network evoluves, whereas others do not. 
The comparison of  both type models would uncover the influences of topological change of the network on cooperation or affluence. 
We present the results of simulations  of the poroposed models in the section 3.
After performing extensive computer simulations, we select and present only the most conclusive results in this article. 
In the last section, we briefly summarize our main findings.   
%The simulation results of all models are too massive, and descriving all results also make essential statements unclear.
%So  we descrive only crucial results which need to draw conclusions in this section. 
%The last section is devoted to summaries. 

 \section{PGG on Network and Models}
 \subsection{PGG}
PGG is a game with two strategies that puts into a public pot (C-strategy, cooperators) or not(D-strategy, defectors) for  public benefit.
 Players secretly choose how many of their private tokens to put into the public pot. 
The tokens in this pot are multiplied by a factor $r$ (synergy factor), where $r$ is normally greater than one and less than the number of players $N$, and the payoff of this public goods is evenly divided among all players. 
 So the tokens are kept even in players who do not contribute. 
The group's total payoff is maximized when everyone contributes all of their tokens to the public pool. 
The Nash equilibrium in this game, however, is not to contribute by all. 
Those who contribute below average or nothing are called "free riders"\cite{Hau}. 

When a player plays a game with $N-1$ players,  we assume that the tokens to put into a public pot are $M>0$. 
So cooperators gain the following payoff $P_c$;
\begin{equation}
P_c=\frac{rmM}{N}-M=M(\frac{rm}{N}-1)
\end{equation}
 where $m$ is the number of cooperators and it is assumed all cooperators contribute the same value $M$. 
 The payoff $P_d$ gained by defectors who contribute nothing is
 \begin{equation}
P_d=M\frac{rm}{N}.
\end{equation}
 So rational players always choose D-strategy because $P_c< P_d$. 
However, all players gain more profit at large $m$ than at small $m$ in general.  
The larger the number of cooperators becomes, the larger payoffs gained by players become.   
That is a dilemma. 
This dilemma is analogous to Prisoners' dilemma where the Paret optimal is not Nash equilibrium. 

\subsection{PGG on Networks}
We extend the original PGG to PGG on networks in this subsection\cite{Santos2}. 
For example, let's consider a network given in Fig.1, where $n$ players exist on nodes. 
Players mutually connected by edges mean that there are some personal relations  among them. 
Nodes connected with a node  by edges represent  opponents who the player on the node plays PGG. 
In Fig.1, player A plays PGG with players B, C, D and E. 
This PGG is called A-chair game, which is played by 5 players.  
On the other hand, player D plays PGG with 2 players, A and E. 
So in Fig.1 the player A  have to play 5 games, that are  B, C, D and E-chair games in addition to A-chair game.  
The total payoff  of the player A is the sum of the 5 games. 
In general, every player i plays  $k_i+1$ PGGs, where $k_i$ is the degree of node i. 
Though a player i contribute to a fixed amount $M$ for all $k_i+1$ PGGs in the previous article \cite{Toyo2},  
we assume that cooperators contribute a fixed amount per member of the group including  $k_i+1$ players, 
\begin{equation}
C_i= \frac{M}{k_i+1}
\end{equation}
in this article. So the payoff acquired by a player i is given by
\begin{equation}
P_i = \sum_{j=1}^{n} B_{ij} \frac{r\sum_{\ell=1}^{n} B_{j\ell} S_\ell C_\ell}{k_j+1} -(k_i+1)S_i C_i, 
\end{equation}
where $B_{ij}$ is defined as follows by using the adjacency matrix $A_{ij}$ and the unit matrix $I=\{I_{ij}\}$
\begin{equation}
B_{ij}=A_{ij}+I_{ij}
\end{equation}
and 
\begin{align*}
S_{i}=
\begin{cases}
1\ \quad( \mbox{for strategy C}    )\\
0\quad( \mbox{for strategy D}).
\end{cases}
\end{align*}

\subsection{Strategy Models}%%%%%%%%%%%%%%%%%%%%2.3%%%%%%%%%%%%%%%%%%%%%
We propose some models for strategy evolution.
They can be broadly classified into the following two types. 
%are divided into two large groups. 
One is that a player playing many PGGs generally employs different strategies for each game, which are represented by  a strategy vector and called A-Model.  
Another  is that a player playing many PGGs employs the same strategy for each game, which is called B-Model. \\

A-Model: Every player i may play $k_i+1$ games with different strategies. 
So every player is characterized by a strategy vector $S^A_i$;
\begin{equation}
S_i^A=( s_{i1}^A, s_{i2}^A, s_{i3}^A, \cdots ,s_{i n}^A), 
\end{equation}
where $s_{ij}^A=\{0,1\}$.
 $s_{i j}^A=1$ when the player does put a token into a public pot and  $s_{ij}^A=0$ when the player 
doesn't. 
The dimension of the vector $S_i^A$ is fixed to the number of players, $n$, so that 
 we keep the dimension of the strategy vectors  for all players at constant value. 
We have to prepare the strategies even for unconnected players, because  the opponents for a player
may change moment by moment due to topology change of the network. 
The strategy vectors for all players are integrated into a matrix; $n \times n$ Strategy Matrix; $\bf S^A$.    
 \\

B-Model: Every player i plays $k_i+1$ games with the same strategies, C or D. 
So the strategies for all players are integrated into a vector $S^B$;
\begin{equation}
S^B=(s_1^B, s_2^B, s_3^B,\cdots ,s_n^B), 
\end{equation}
where $s_i^B=\{0,1\}$.
 $s_{i}^B=1$ when the player does put a token into a public pot and  $s_{i}^B=0$ when the player 
doesn't.   

\begin{figure}[t]
\centering
	\includegraphics[width = 6.0cm]{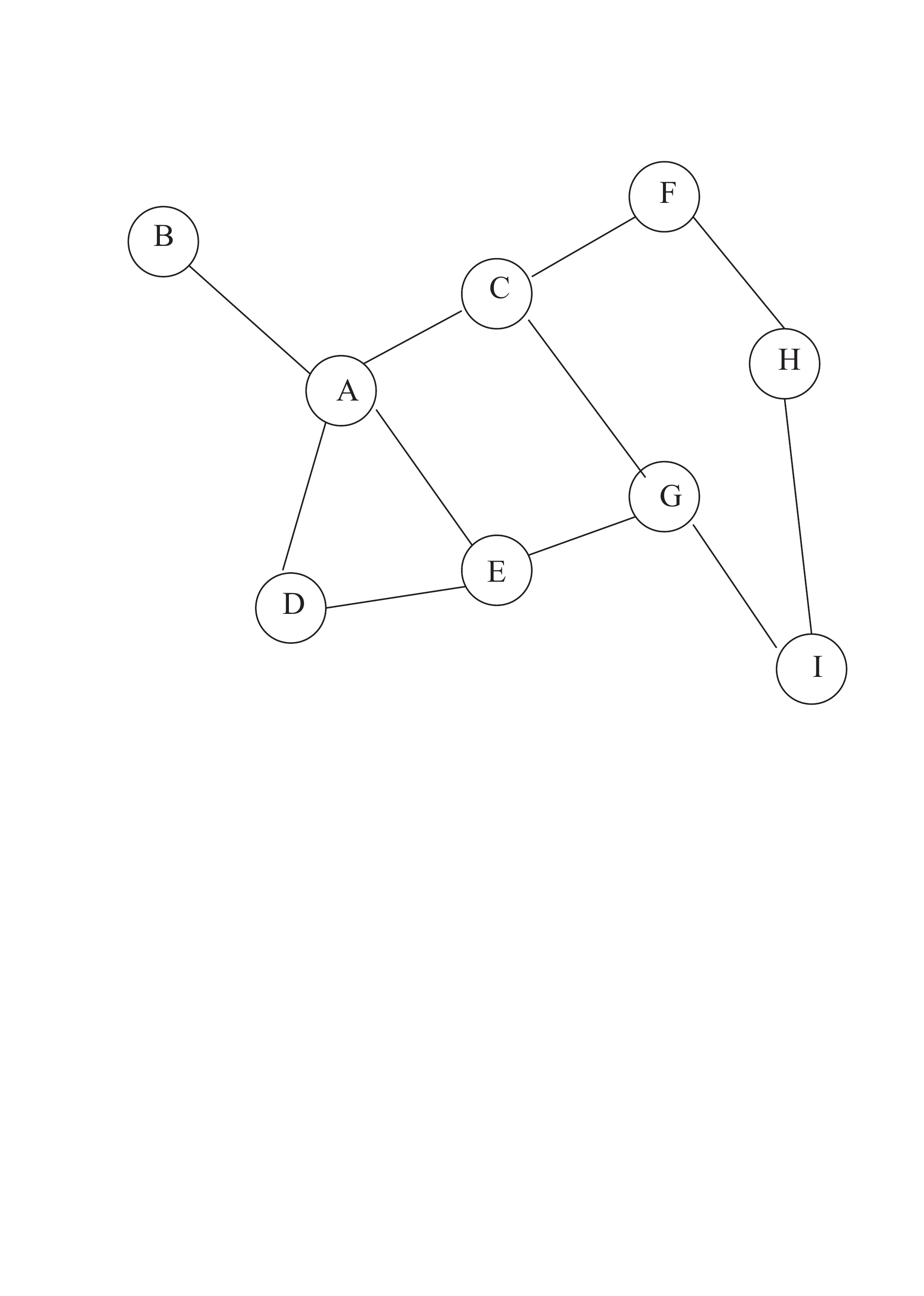}
	\caption{PGB on Network .}
\end{figure}

 Initially we take $s_i^A=1$ or $s_i^B=1$ with the probability $c$ that is a model parameter and $s_i^A=0$ or $s_i^B=0$ with the probability $1-c$. 
Some updating models for strategies are obtained for A-model and B-model, respectively. 
We introduce two models for tactics to explore the effects of the game dynamics for A-Model. \\

A-1 Model is the model that a player i imitates the strategy vector of the player who acquired the largest payoff
among players connected with the player i in the next round. 
Thus the strategies of players are affected by the game dynamics. 
They are greedy persons.  \\

A-2 Model is the model that a player i increases the ratio of C(D) by $t$ in his/her strategy vector, 
when the total number of C(D) strategy
in i-$th$ component of the strategy vectors of the players connected with the player i is bigger than the one of D(C), 
 in the next round. 
Thus the strategies of players are not affected by the game dynamics. 
Then it is expected to obtain the same results as the previous studies \cite{Toyo2}. 
They are opportunism. \\ 

Similarly to A-Model, three models for tactics are introduced in B-Model. \\

B-1 Model is the model that a player i imitates the strategy of the player 
that acquired the largest payoff among players connected with the player i in the next round. 
Thus the strategies of players are affected by the game dynamics like A-1 Model. 
Then it is expected to obtain the same results as the previous studies \cite{Toyo2}.  
They are also greedy persons.  \\

B-2 Model is the model that a player i chooses the strategy of the majority side among 
the  players connected with the player i, in the next round. 
Thus the strategies of players are not affected by the game dynamics like A-2 Model. 
They are also opportunism. \\

B-3 Model is the model that a player flips his/her strategy from C(D) to D(C) in the next round, when the payoff acquired by the player becomes negative. 
They are  persons who hate to admit defeat.
%This model is more slight restraint than the model proposed in \cite{Ebel2} where if a player get a higher payoff, the player  aggressively flips the player's strategy. 
Then the players with D-strategy basically acquire positive payoffs, since they are free riders. 
So it is considered that there are the cases that C-strategy change to D-strategy but there are not the opposite pattern. 
 It is expected that D-strategy increases. \\

We do not consider any model corresponding to B-3 Model in A-Model, because we think that a person 
 would not drastically change his(her) strategy 
such as flipping all the components of a strategy vector in a practical manner. 
Such thought has already seen in A-2 Model where the change of the ratio of C or D strategy in a strategy vector is controlled by a not so large parameter $t$.

\subsection{Topology Change  Models}%%%%%%%%%%%%%%%%%%%%%%%%%%%%%%%%%%%%%%%%%
In the early days, a model for Prisoners' Dilemma on random networks has been proposed where when the payoff of a focal player increases, the player accepts a randomly chosen new link   and when the payoff decreases, the player cut a randomly chosen link in \cite{Ebel}.  
The topology change in this model induces C-C links and  suppresses D-D links. 
Moreover a model that is implicitly based on payoffs, in which a player who the strategy is mimicked by another player make another different friend has been proposed\cite{Perc2}. 
It is also shown that avoiding connections with players with D-strategy often promote cooperation even at lower $r$\cite{Wu1}.
Furthermore the preferential selection for topology change model has been also explored\cite{Shi}. 
The way of the topology change in the article is based on strategy-inspired\cite{ZZhang} and  includes implicitly phycological factor as shown below.  
  We proposed two models for topological evolutions of a network\cite{Toyo2}.  
Topology change of our model is not directly based on increase and decrease of the payoff but is to rupture a player who is not favored as shown below. 
This can be thought to be a natural way.

 %%%%%%%%%%%%%%%%%%%%%%%%%%%%%%%%%%%%%%%%%%%%%%%%%%%%%%%%%%%%%%%%%
%%%%%%%%%%%%%%%%%%%%%%%%%%%%%%%%%%%%%%%%%%%%%%%%%%%%%%%%%%%%%%%%

 $\alpha$-$Model$\\
 A player i cuts the edge with one of players with D strategy, who are harmful to the player i or reduce the payoff of  the player i,  with the probability $p$.  
 The player i sets up  a new edge to an unconnected player j chosen randomly with the probability $q$\cite{Fu} inspired the models 
introduced in \cite{Santos1}, \cite{Zim2}, \cite{Wu1},\cite{Van}. 
However, our model is so simple that the time scale, which has been introduced in most of the above studies,  with respect to coevolution is not introduced. 
We attempt to realize cooperation without the time scale in this article.

$\beta$-$Model$	\\
A player i cuts an edge with one of players with D  strategy, which are harmful to the player i or reduce the payoff of  the player i,  with the probability $p$.  
 The player i sets up  a new edge to an unconnected player j who is not directly connected but connected in two hops with the player i  with the probability $q$\cite{Zim2}. \\
 
 When a player connect an edge to some new friend introduced by our friends, the concept such the reputation\cite{Fu21} is not used,  since  we pursue preferably simple models for cooperation. 
A property of the two models for topology change is that players break off their edges with players employing D-strategy that is unprofitable for the players. 
This rule for the topology change may create a potential for promoting cooperation. 
We study this possibility. 

When all edges from a node is broken (it does not mean that the node breaks  an edge itself that connects the node to another node) and the node is isolated, it is impossible that the node is connected with any nodes again in $\beta$ model. 
Thus the number of isolated nodes increase over time. 
By contrast, a node that all edges from the node is broken and is isolated may be accidentally connected with some nodes  in $\alpha$ model. 
So the isolation phenomenon will be relieved. 
 We, actually, study only $\alpha$ model in this article and will study $\beta$ model in another time.  
 
We make simulations of  these models on the following typical three networks as initial networks to explore 
the transition of the average payoff and the number of cooperators, and topological transition, especially the degree distributions for all models. 

1. Random Networks(ER, Erdos-Reny model\cite{Ren}), 

2. Scale Free networks (SF, Barabashi-Albata model\cite{Bara}) with scaling exponent=3, 

3. Small World networks(WS, Wattz-Strogatz models\cite{ws} with rewriting ratio $w=0.1 \sim 0.01$), 
 
where $w$=0.01 $\sim$ 0.1 is the parameter area where networks show obvious small world properties.

%%%%%%%%%%%%%%%%%%%%%%%%%%%%%%%%%%%%%%%%%%%%%%%%%%%%%%%%%%%
\section{Simulation Results}%%%%%%%%%%%%%%%%%%%%%%%%%%%%and $c=0.5$
In simulations, we fix $t=0.1$  and take $n=200$ as  the network size. 
For simplicity, we set $M=1$ but this does not bring crucial effects on the results of this article. @
The average degree $k$ is taken between $4$ and $16$. 
$k=4 \sim 6$ is representatives of low degrees and $k=16$ is a representative of high degrees
 for $n=200$. 
 $r$ is taken between $0.5$ and $12.0$. 
 Though $r=0.5$ seems to be trivial since it is expected that all D-strategy is realize, simulations do not necessarily follow in the intricate setting given in these coevolutionary models.  
 In this section, varying $p$, $q$, $r$, $k$ and $c$, we investigate the ratio of D-strategy, the average payoff over all players and the future of the constructed network topology after topology change, especially degree distributions. 
$c$ has only an influence on the final converged value of D-ratio but does not have any influences on the 
whole behaviors of the time change of the D-ratio. 
So we fix $c=0.5$. 
When  $0<\{p, q\}<1$, the results are essentially the same as those of $p=q=1 $, but only the effect taking $0<\{p, q\}<1$ is that  convergence  becomes slow. 
So we should study the cases with $p=q=0$ and $p=q \neq 1 $ to explore the influence of topology change on the  average payoffs and the number of cooperators. 
At $p=q$, the total number of edges is kept nearly constant. 
These different two values of $p$ and $q$ must reveal the effects of topology change.  
%$k$ is taken between $4$ and $16$. 
%$k=4 \sim 6$ is representatives of low degrees and $k=16$ is a representative of high degrees
 %for $N=200$. 
 %$r$ is taken between $0.5$ and $8.0$. 
 %Though $r=0.5$ seems to be trivial since it is expected that all D-strategy is realize, simulations do not always show such results.  
The number of repetition is appropriately chosen in the same parameters by observing
 how the results converge.

%In view of such properties, we show mainly the chracteristic results of $\alpha$-model in this article. 
%%%%%%%%%%%%%%%%%%%%%%%%%%%%%%%%%%%%%%%%%%%%%%%%%%%%%%%

\subsection{A models}
	
\textbf{A-1model:} 
The ratio of C-strategy is almost same as the one of D-strategy in the cases without topology change, but 
it is observed that C-strategy  outnumber D-strategy a little  in WS networks with small $k$ at large $r$ as shown in Fig.2 and Fig.3.(f1 f2)
This property is like the previous model that a player contributes to a fixed amount \cite{Toyo2}. 
However, the model with topology change on WS networks and ER networks clearly promotes C-strategy  irrespective of $r$, than the previous model,. 
Fig.4 and Fig.5 show the ratio of D-strategy in WS-nets, but the ones in ER-nets are essentially the same as these figures. ( figf3f4) 
SF networks do not so, especially at small $r$ which are shown Fig.6 and Fig.7. ( f5f5') %%%%%%%%%%%%
This behavior is like the cases without topology change and also like the previous model that a player contribute to a fixed amount \cite{Toyo2}. 
The inhomogeneity of the degree in initial networks is thought to resist the increment of C-strategy, especially at small $r$. 
We should note that the final degree distribution that will be discussed in the subsection 3.3 are Poisson like degree distribution against the initial one.  
These results stand in contrast to the previous model where D-strategy is almost dominant in the case without topology change and C-strategy is sometimes promoted only at large $r$ in the case with topology change.   

It may be suggested that the synergism between the topology change and the game dynamics promotes C-strategy from all of the results.  
By studying  A-2 model, we think that the effect of topology change works more intensely than the game dynamics.  

Regardless of whether with or without the topology change, the average payoff in WS and ER networks behaves in the  same way as shown in Fig.8 and Fig.9, (f6 f7) where the average payoffs are negative only at $r$=0.5.
This is a natural result. 
In SF network, however, the average payoffs are always positive as shown in Fig.10 and Fig.11. f8f8'
This shows that the average payoff does not become negative even in the cases where there are a lot of D-strategy.

\begin{figure}[tbp]
 \begin{minipage}{0.45\hsize}
 \begin{center}
\includegraphics[width = 7.0cm,height=3.5cm,clip]{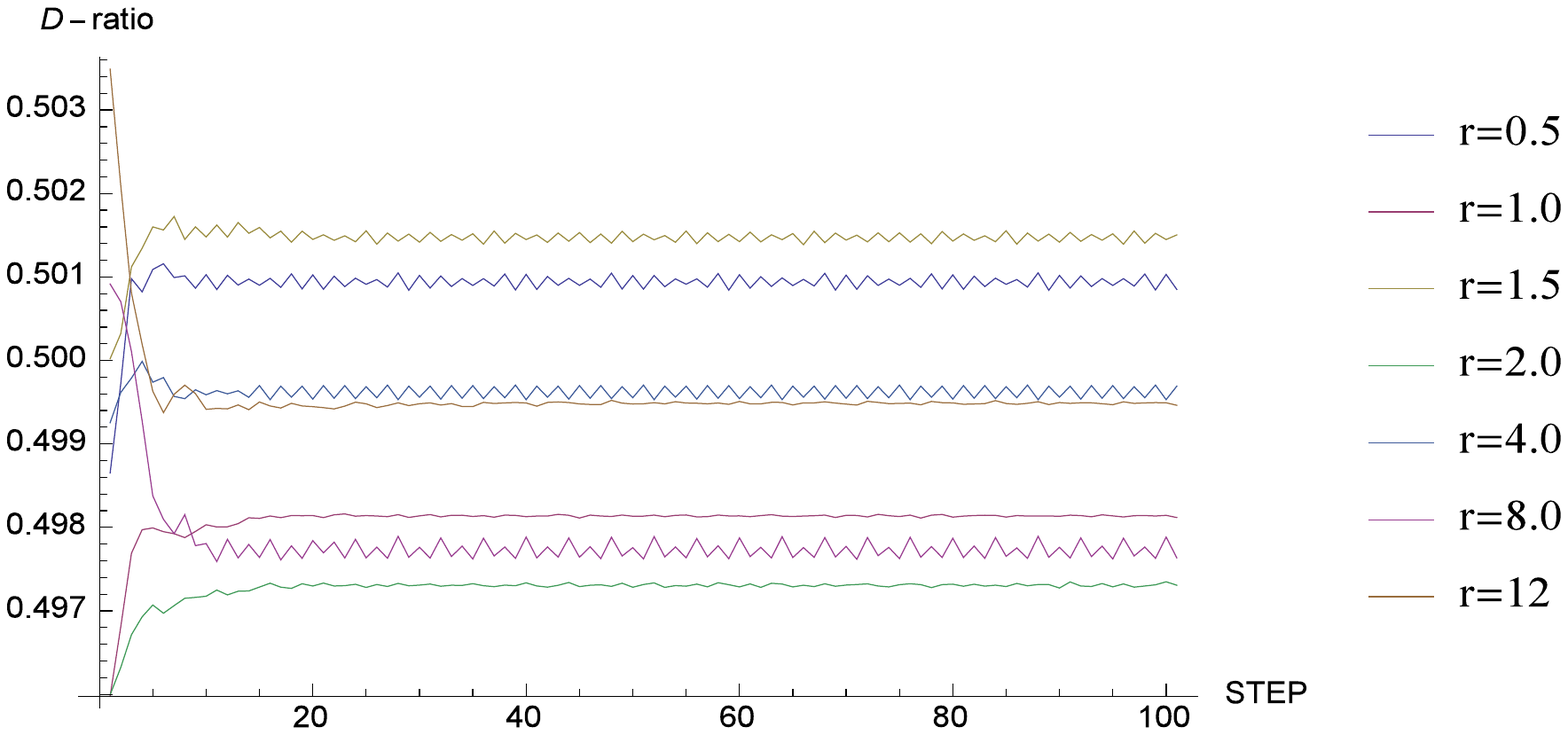}
 \end{center}
 \hspace*{3mm}
 \caption{\small The D ratio in A1-model (ERnet 
  $k=4$)  without topology change.}
\label{fig:two}
\end{minipage}
\hspace*{3mm}
\begin{minipage}{0.45\hsize}
 \begin{center}
\includegraphics[width =7.0cm,height=3.0cm,clip]{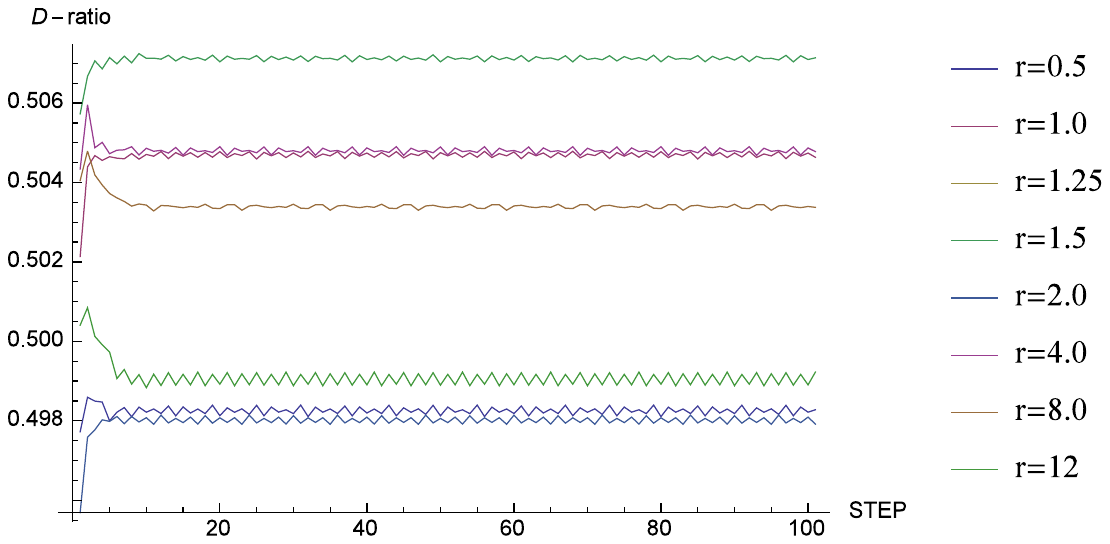}
 \end{center}
 \caption{\small The D ratio in A1-model (WSnet 
$w=0.01$ and $k=4$) without topology change.}
 \label{fig:three}
\end{minipage}
\end{figure}
%%%%%%%%%%%%%%%%%%%%%%%%%%%%%%
 \begin{figure}[tbp]%[!hbt]
 \begin{minipage}{0.5\hsize}
  \begin{center}
\includegraphics[width = 7.0cm,height=3.5cm,clip]{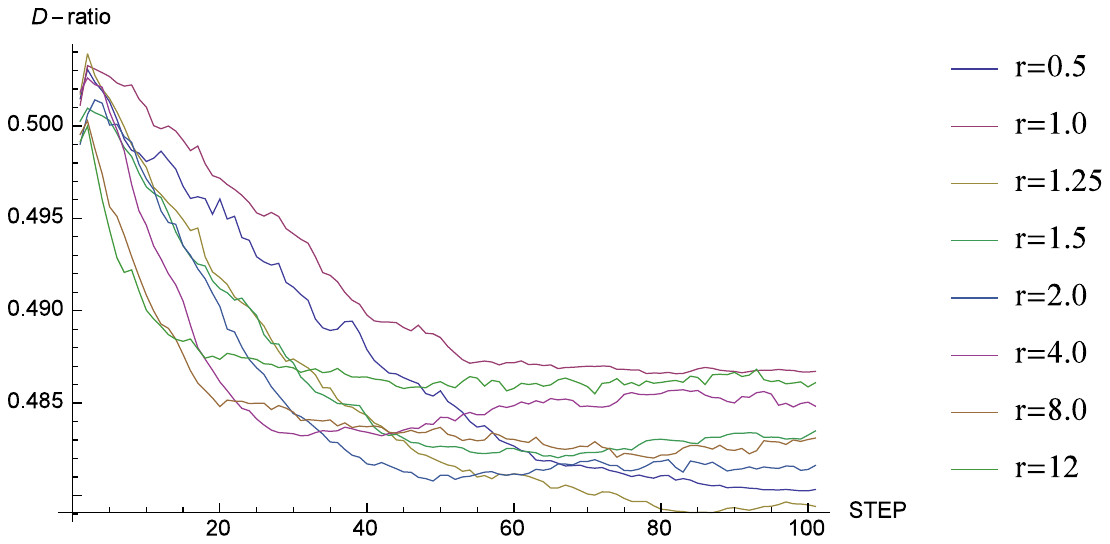}
 \end{center}
 \caption{\small The D ratio in A1-model (WSnet 
$w=0.01$ and $k=4$) with topology change.}
\label{fig:four}
 \end{minipage}
 \hspace*{3mm}
\begin{minipage}{0.5\hsize}
\begin{center}
\includegraphics[width = 7.0cm,height=3.5cm,clip]{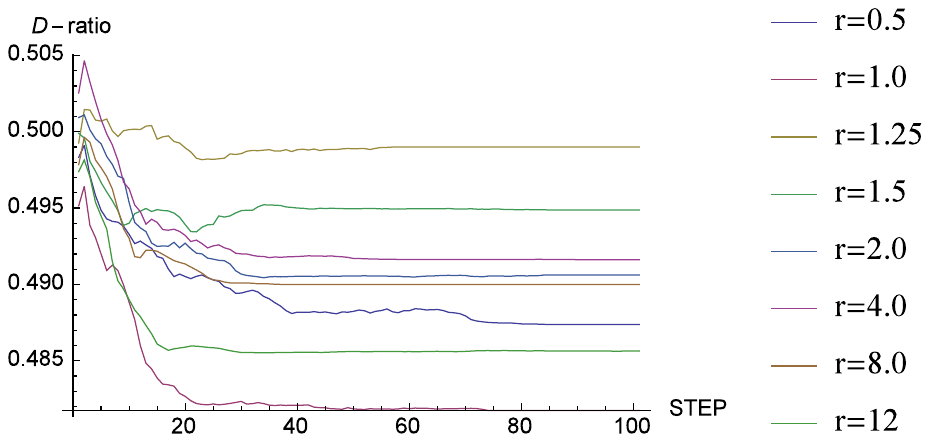}
\end{center}
\caption{\small The D ratio in A1-model (WSnet 
$w=0.01$ and $k=16$) with topology change.}
\label{fig:two}
\end{minipage}
\end{figure}
%%%%%%%%%%%%%%%%%%%%%%%%%%%%%%

%%%%%%%%%%%%%%%%%%%%%%%%%%%A1 %%%
 \begin{figure}[tbp]%[!hbt]
 \begin{minipage}{0.5\hsize}
  \begin{center}
\includegraphics[width = 7.0cm,height=3.0cm,clip]{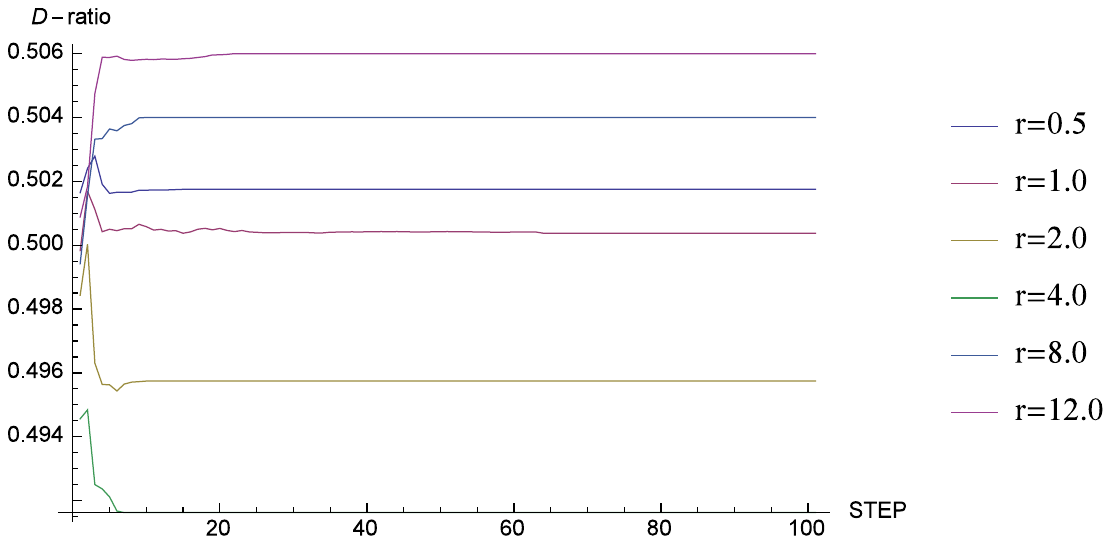}
 \end{center}
 \caption{\small The D ratio in A1-model (SFnet 
 $k=16$) with topology change.}
\label{fig:four}
 \end{minipage}
 \hspace*{3mm}
\begin{minipage}{0.5\hsize}
\begin{center}
\includegraphics[width = 7.0cm,height=3.0cm,clip]{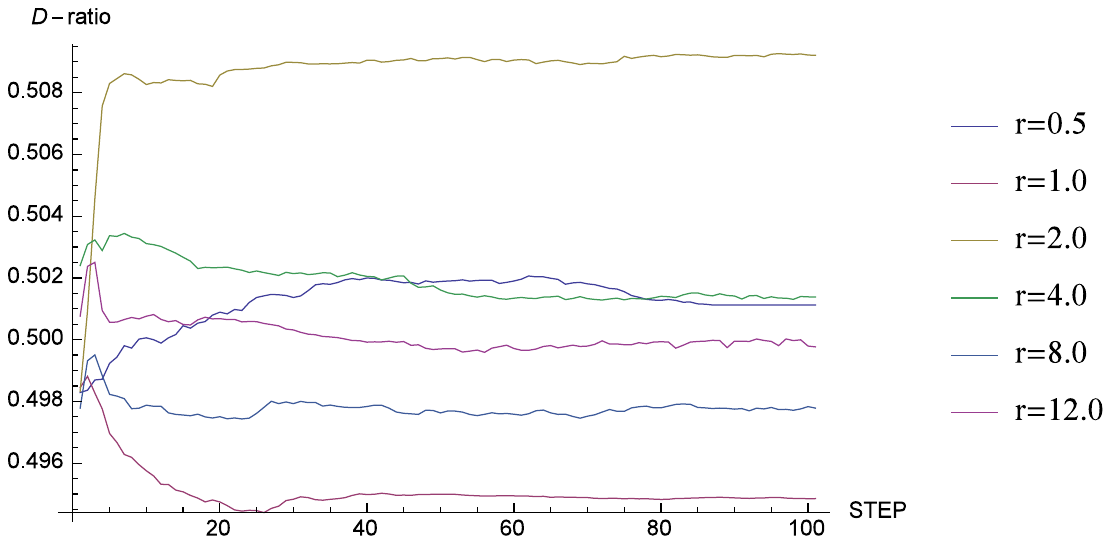}
\end{center}
\caption{\small The D ratio in A1-model (SFnet 
 $k=6$) with topology change.}
\label{fig:two}
\end{minipage}
\end{figure}
%%%%%%%%%%%%%%%%%%%%%%%%%%%%%%
%%%%%%%%%%%%%%%%%%%%%%%%%%%%%%payoff

\begin{figure}[tbp]
 \begin{minipage}{0.5\hsize}
  \begin{center}
\includegraphics[width = 7.0cm,height=3.5cm,clip]{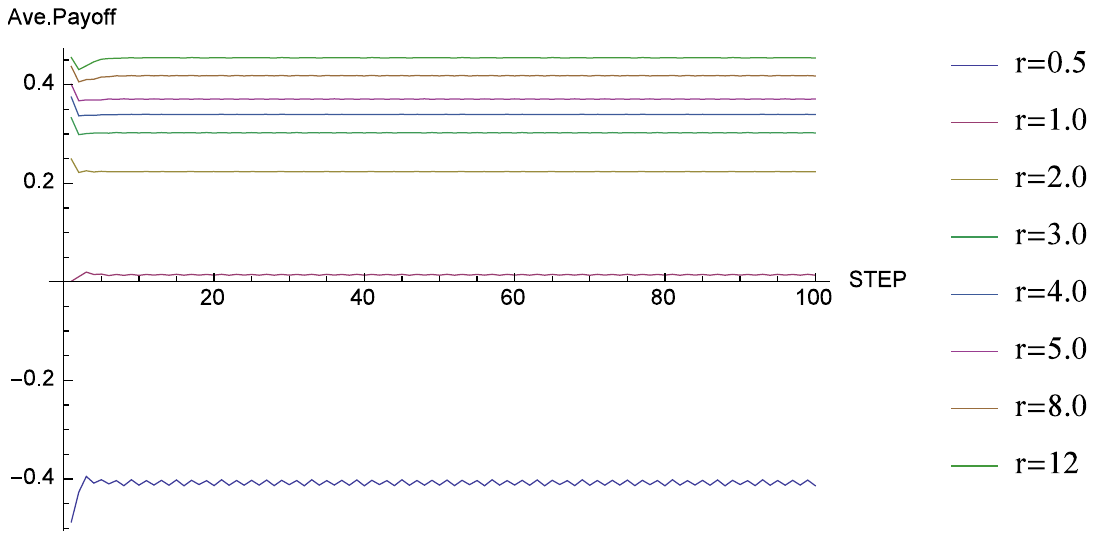}
 \end{center}
 \caption{\small The average payoff in A1-model (WSnet 
 $w=0.01$ and $k=4$) without topology change.}
\label{fig:four}
 \end{minipage}
 \hspace*{3mm}
\begin{minipage}{0.5\hsize}
\begin{center}
\includegraphics[width = 7.0cm,height=3.5cm,clip]{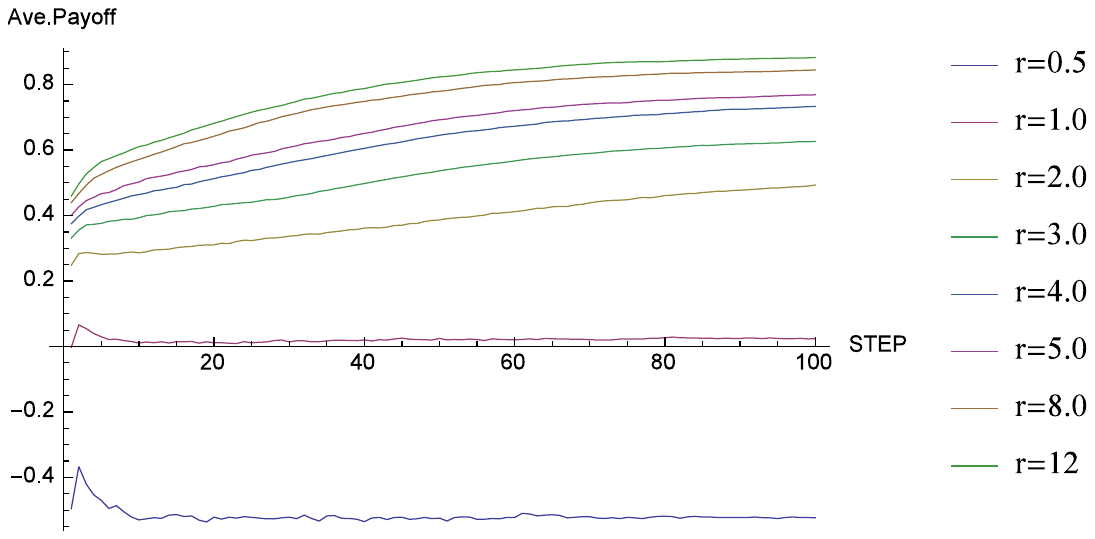}
\end{center}
\caption{\small The average payoff in A1-model (WSnet 
 $w=0.01$ and $k=4$) with topology change..}
\label{fig:two}
\end{minipage}
\end{figure}

\begin{figure}[tbp]
 \begin{minipage}{0.5\hsize}
  \begin{center}
\includegraphics[width = 7.0cm,height=3.5cm,clip]{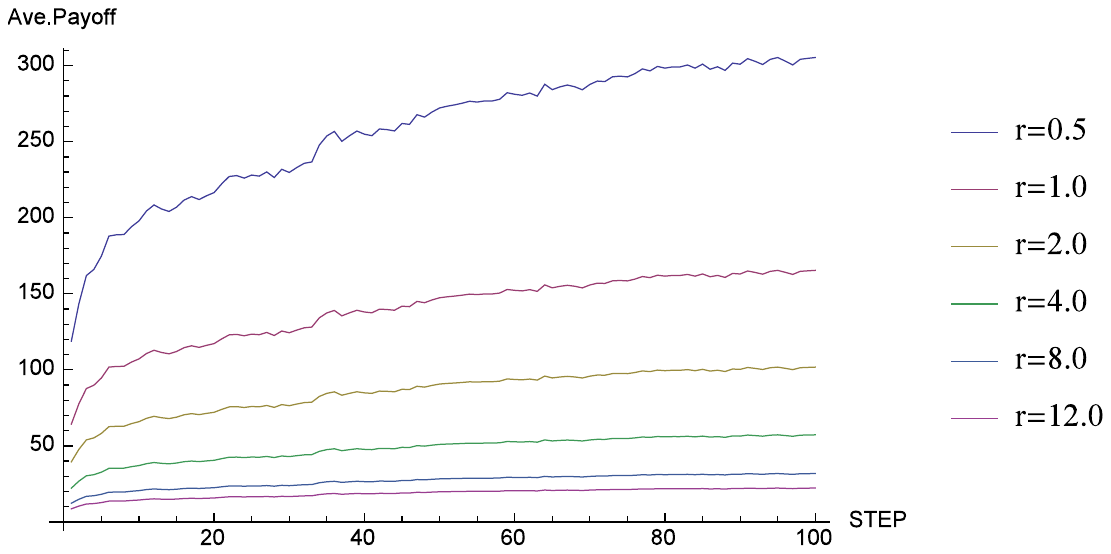}
 \end{center}
 \caption{\small The average payoff in A1-model (SFnet 
  $k=16$) with topology change.}
\label{fig:four}
 \end{minipage}
 \hspace*{3mm}
\begin{minipage}{0.5\hsize}
\begin{center}
\includegraphics[width = 7.0cm,height=3.5cm,clip]{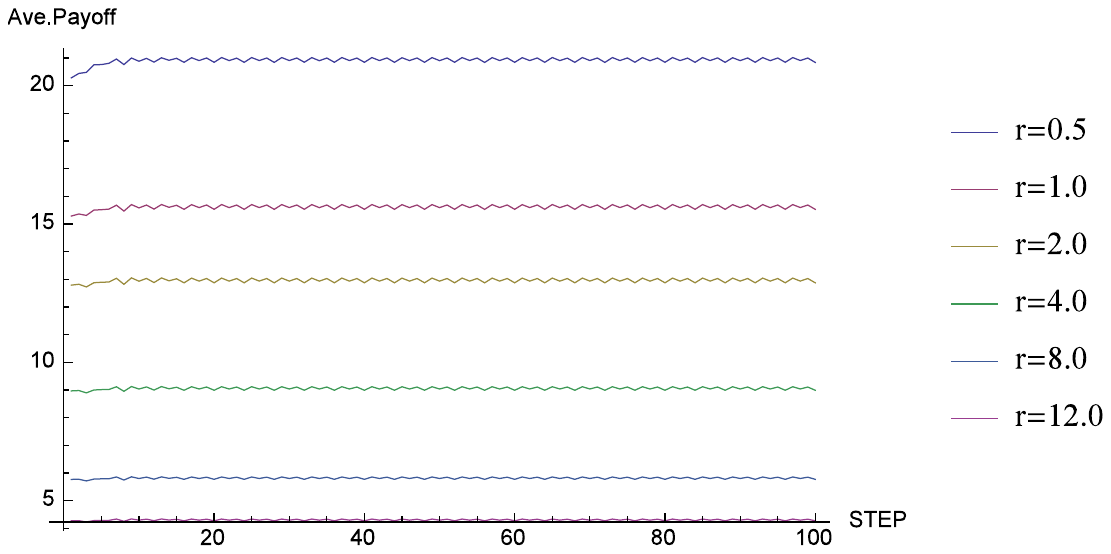}
\end{center}
\caption{\small The average payoff in A1-model (SFnet 
 $k=6$) without topology change..}
\label{fig:two}
\end{minipage}
\end{figure}
%%%%%%%%%%%%%%%%%%%%%%%%%%%%%%

 %%%%%%%%%%%A3 models
\textbf{A-2model:}
The results are essentially the same as those of the previous models\cite{Toyo2}. 
 This is because A-2 model does not depend on the game dynamics but the studies are refined in this article. 
While D-strategy is more or less  superior to C-strategy in the cases without topology change at small $k$, 
the ratio of C-strategy depends on $r$ at large $k$.  
They are represented in Fig.12-Fig.15.  f9f10,(f11,f12). 
As for  the cases with topology change, C-strategy, however,  is remarkably superior to D-strategy at small $k$  which are shown in Fig.16 and Fig.17. f13f14  
 This trend is relieved at large $k$ as shown in Fig.18 and Fig.19. (fig15f16)
 
 The average payoff in all networks  is positive except for the case of $r$=0.5 even in SF networks like A-1 model.
 We can see that topology change promotes C-strategy in this model.
 This model suggests the possibility that topology change only promotes C-strategy even without game dynamics.

%These models with topology change and small $k$ promote C-strategy in all initiial networks as shown in Fig.7. 
%Then C strategy is promoted in almost cases, even the models without topology change, at small $k$ and large $r$ as shown in Fig.8. %%%%%%%%%%%%%%%%
%Only when only $k$ is small and $r$ is large, the cases without topology change C strategy is also promoted, 
%But D-strategy is promoted for the cases with large $k$ in general, regardless of $r$.  
%It is partly observed in Fig.9. 
%Thus  C-strategy is clearly promoted more in the cases with topology change than the ones without topology change.   
%So topology change promotes C-strategy when players go along with the strategies of players surrounding one. 
%Since this model is independently of game dynamics, the promotion of C-strategy is not due to an interaction between dynamics and topology. 
%The average payoff over all players is positive at $r>1$ in the all cases, even in D dominant cases as shown in Fig. 10 andFig.11. 

%Since A-2 model does not depend on game dynamics, the results are the same as the previous studies \cite{Toyo2}.
%Really the same features were shown in the previous model, but here they are refined in this article. 

%%%%%%%%%%%%%%%%%%%%%%%%%%%%%%%%%%%

%%%%%%%%%%%%%%%%%%%%%%%%%%%%%%%%%A2

%%%%%%%%%%%%%%%%%%%%%%B1

 \begin{figure}[tbp]
 \begin{minipage}{0.5\hsize}
  \begin{center}
\includegraphics[width = 7.0cm,height=3.5cm,clip]{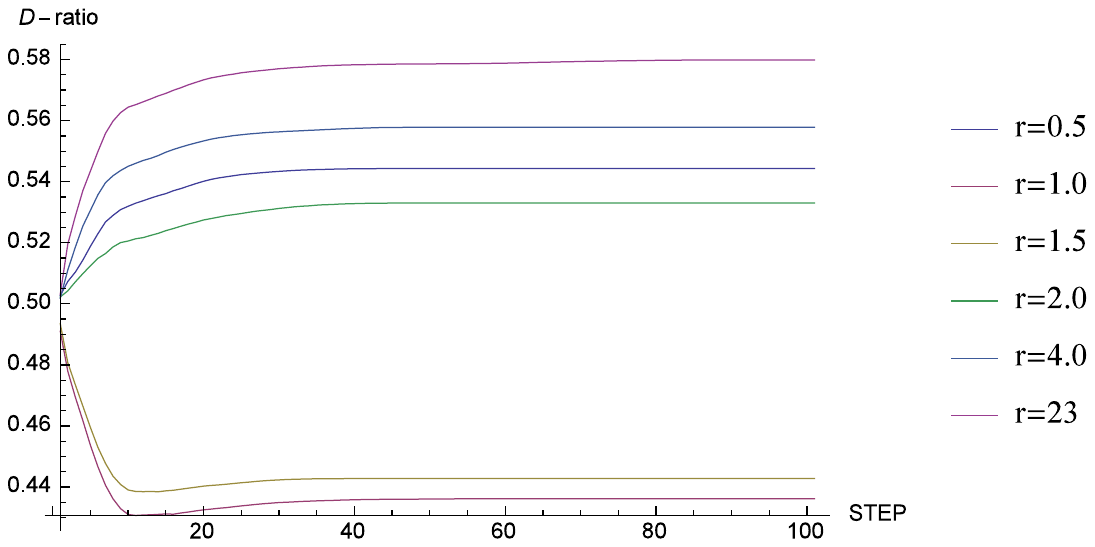}
 \end{center}
 \caption{\small The D ratio in A2-model (WSnet 
 $w=0.1$ and $k=4$) without topology change.}
\label{fig:four}
 \end{minipage}
 \hspace*{3mm}
\begin{minipage}{0.5\hsize}
\begin{center}
\includegraphics[width = 7.0cm,height=3.5cm,clip]{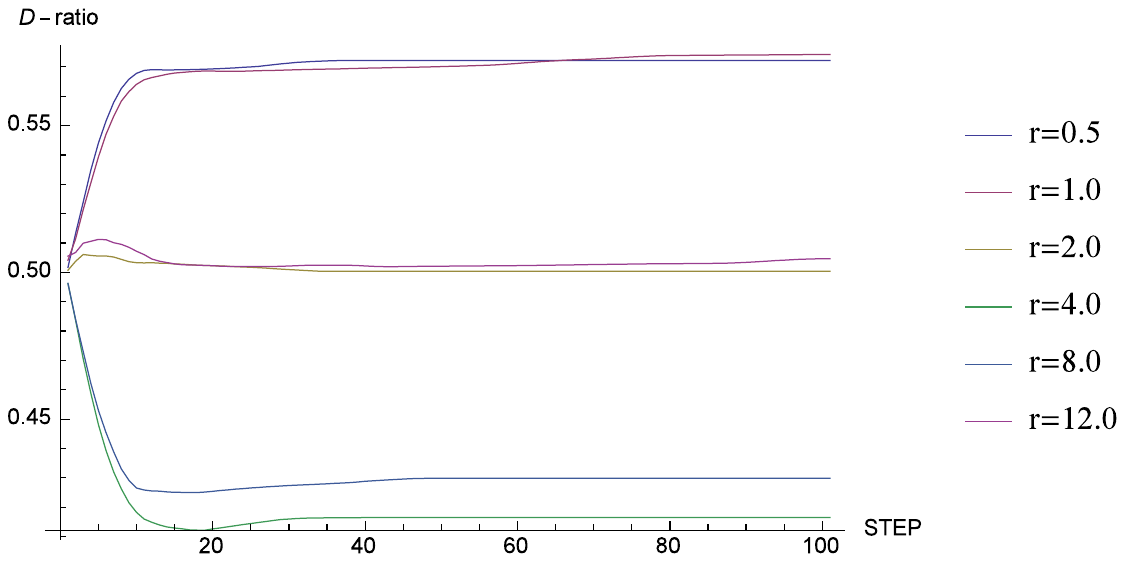}
\end{center}
\caption{\small The D ratio in A2-model (WSnet 
 $w=0.1$ and $k=16$) without topology change.}
\label{fig:two}
\end{minipage}
\end{figure}
%%%%%%%%%%%%%%%%%%%%%%%%%%%%%%A3

%%%%%%%%%%%%%%A3%%%%%%%%%%%%
 \begin{figure}[tbp]
 \begin{minipage}{0.5\hsize}
  \begin{center}
\includegraphics[width = 7.0cm,height=3.5cm,clip]{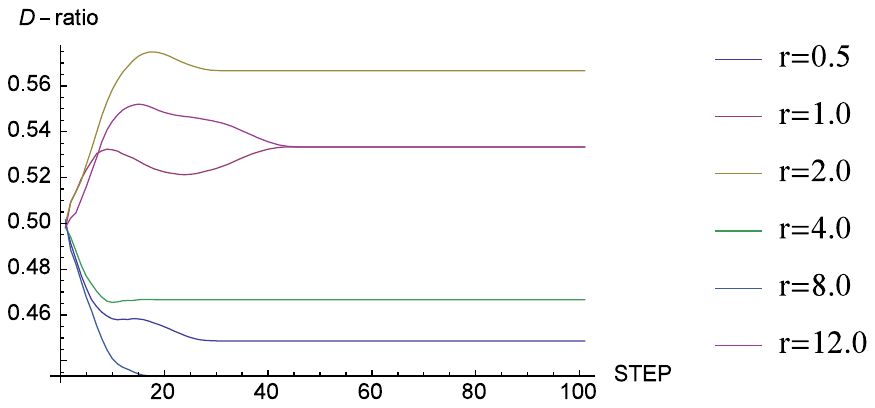}
 \end{center}
 \caption{\small The D ratio in A2-model (ERnet 
 $k=16$) without topology change.}
\label{fig:four}
 \end{minipage}
 \hspace*{3mm}
\begin{minipage}{0.5\hsize}
\begin{center}
\includegraphics[width = 7.0cm,height=3.5cm,clip]{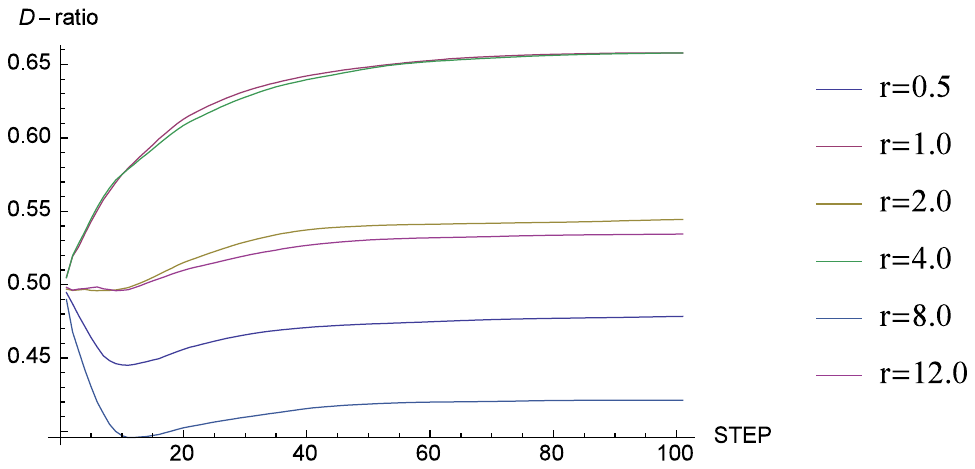}
\end{center}
\caption{\small The D ratio in A2-model (ERnet 
 $k=4$) without topology change.}
\label{fig:two}
\end{minipage}
\end{figure}

 \begin{figure}[tbp]
 \begin{minipage}{0.5\hsize}
  \begin{center}
\includegraphics[width = 7.0cm,height=3.5cm,clip]{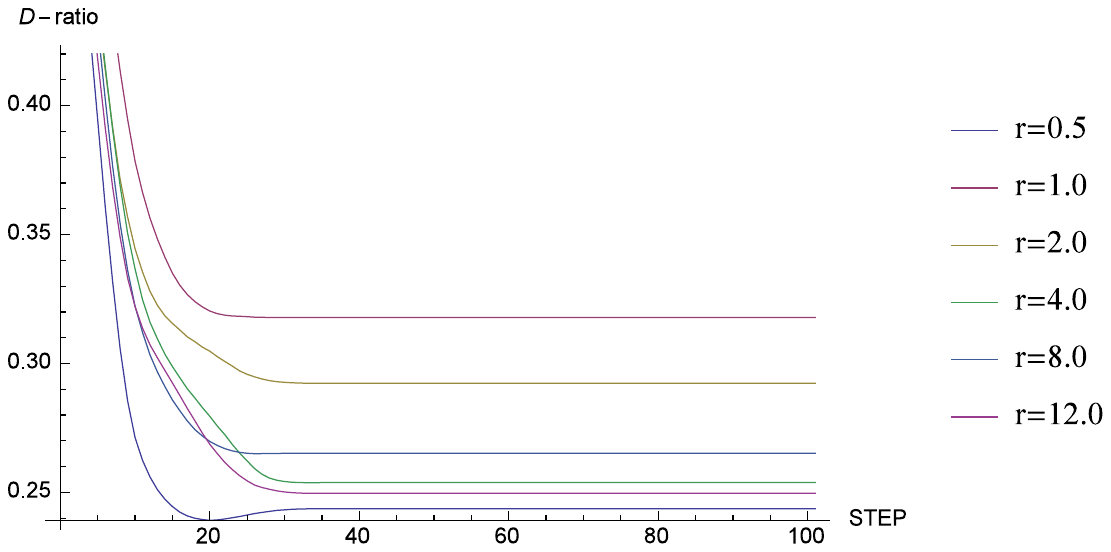}
 \end{center}
 \caption{\small The D ratio in A2-model (WSnet 
$w=0.1$ and  $k=4$) with topology change.}
\label{fig:four}
 \end{minipage}
 \hspace*{3mm}
\begin{minipage}{0.5\hsize}
\begin{center}
\includegraphics[width = 7.0cm,height=3.5cm,clip]{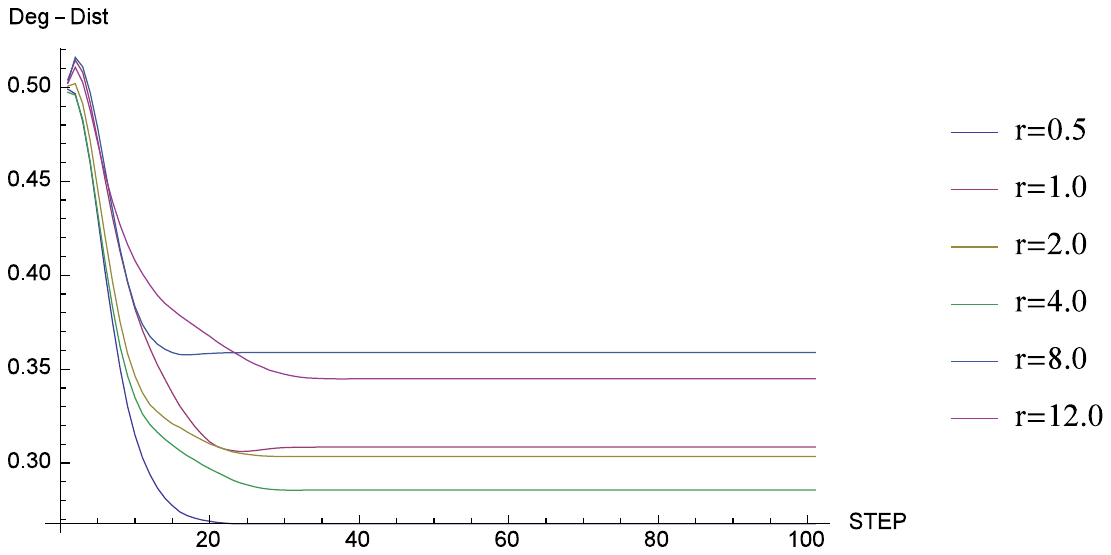}
\end{center}
\caption{\small The D ratio in A2-model (ERnet 
 $k=4$) with topology change.}
\label{fig:two}
\end{minipage}
\end{figure}

 \begin{figure}[tbp]
 \begin{minipage}{0.5\hsize}
  \begin{center}
\includegraphics[width = 7.0cm,height=3.5cm,clip]{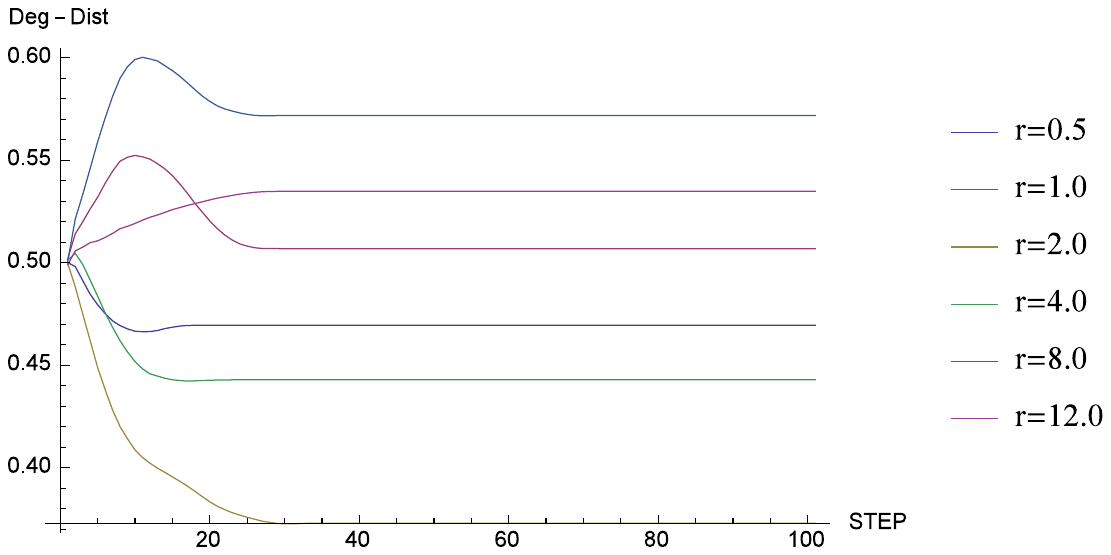}
 \end{center}
 \caption{\small The D ratio in A2-model (WSnet 
 $w=0.01$ and $k=16$) with topology change.}
\label{fig:four}
 \end{minipage}
 \hspace*{3mm}
\begin{minipage}{0.5\hsize}
\begin{center}
\includegraphics[width = 7.0cm,height=3.5cm,clip]{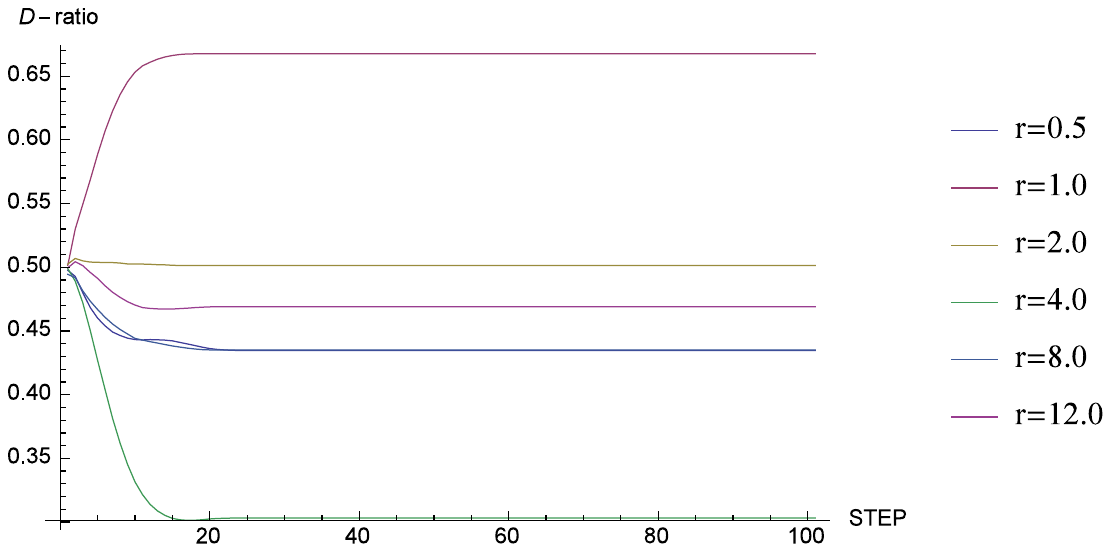}
\end{center}
\caption{\small The D ratio in A2-model (ERnet 
 $k=16$) with topology change.}
\label{fig:two}
\end{minipage}
\end{figure}

%%%%%%%%%%%%%%%%%%%%%%%%%%%%%%B1
\subsection{B models}
\textbf{B-1model:} 
While at small $k$, C-strategy is excessively promoted at large $r$ in the cases without topology change,  
D-strategy mainly is promoted except at large $k$  for large $r$, similar to the previous models\cite{Toyo2}. 
These are represented in Fig.20 and Fig.23.(fig20-23) 
This behavior does not depend on the initial networks.  
This tendency becomes more notable in the cases with topology change as shown in Fig.24 and Fig.25. (24,25)
C-strategy  becomes sometimes predominant over D-strategy at all $r$ for small $k$ as shown Fig.26. (26)
Moreover there are sometimes the situation that D-strategy becomes predominant at all $r$ for large $k$ as shown Fig.27.  (27)
This stands in contrast to the previous models\cite{Toyo2} where D-strategy often accounts for $100$  percent for all $r$ values at large $k$. 
So the power to promote C-strategy is considered to work in the model in this article. 

As for the average payoff,  in almost cases, irrespective of with and without topology change, 
all  average payoffs are positive over all $r$ values as shown Fig.28. f28
The only  exception arises in ER networks with and without topology change at $r < 1$ for small $k$ which is given in Fig.29.     f29
Then  players of about $70$ percent follow C-strategy in ER networks at $r < 1$ and $k=4$. 
Some players adopt D-strategy and they  exploit the players employing C-strategy.   
The average payoff becomes negative due to the players employing C-strategy  of about $70$ percent. 

As a whole, we can not observe  the effect of topology change and so the results mainly depend on the game dynamics. 
Generally C-strategy is promoted at large $r$ and small $k$, and the promotion of C-strategy proceeds even more than the previous models\cite{Toyo2}.

 \begin{figure}[bpt]
 \begin{minipage}{0.5\hsize}
  \begin{center}
\includegraphics[width = 7.0cm,height=3.5cm,clip]{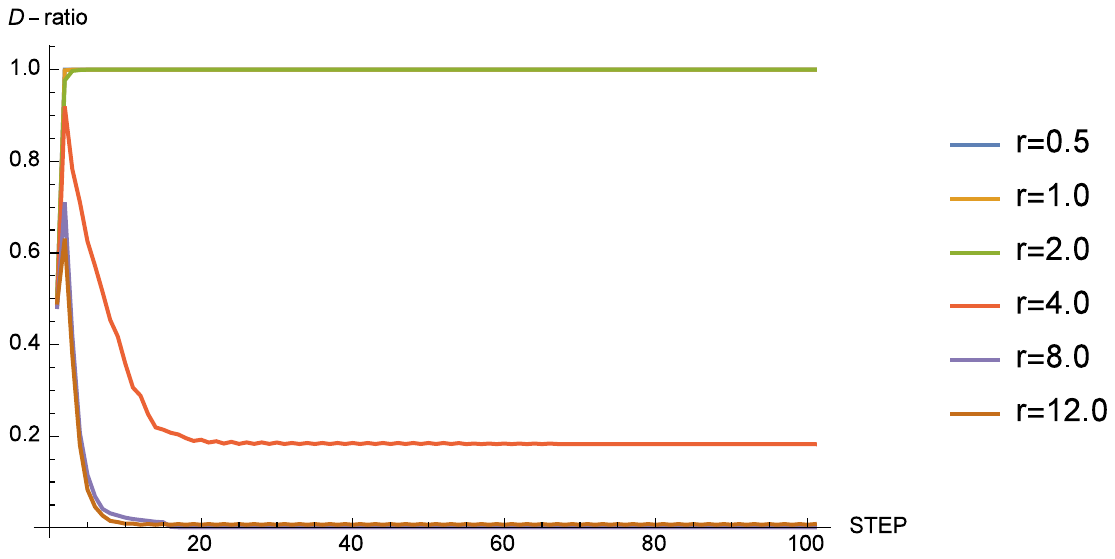}
 \end{center}
 \caption{\small The D ratio in B1-model (WSnet 
 $w=0.01$ and $k=4$) without topology change.}
\label{fig:four}
 \end{minipage}
 \hspace*{3mm}
\begin{minipage}{0.5\hsize}
\begin{center}
\includegraphics[width = 7.0cm,height=3.5cm,clip]{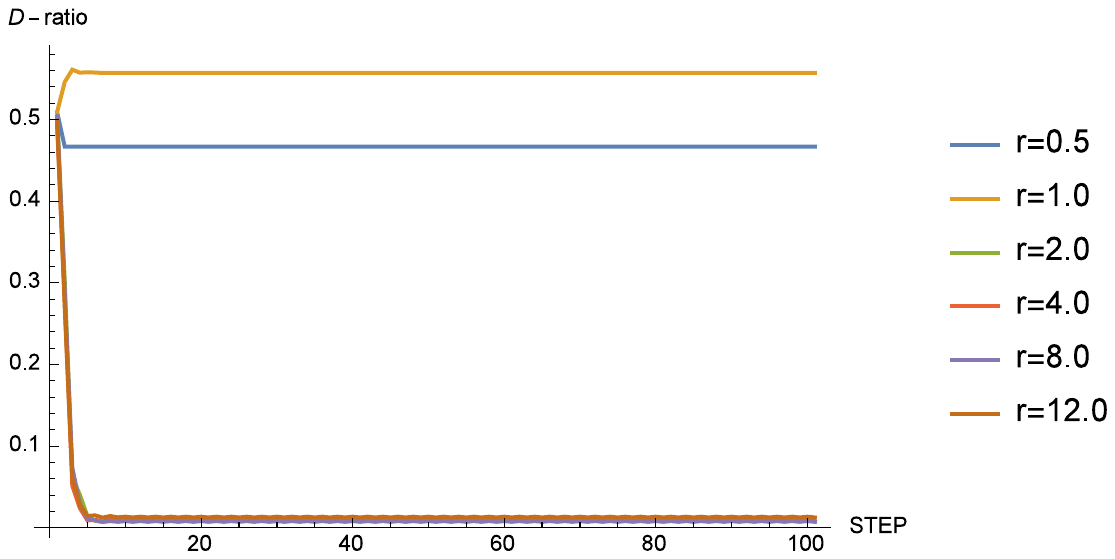}
\end{center}
\caption{\small The D ratio in B1-model (ERnet 
 $k=4$) without topology change.}
\label{fig:two}
\end{minipage}
\end{figure}

 \begin{figure}[tbp]
 \begin{minipage}{0.5\hsize}
  \begin{center}
\includegraphics[width = 7.0cm,height=3.5cm,clip]{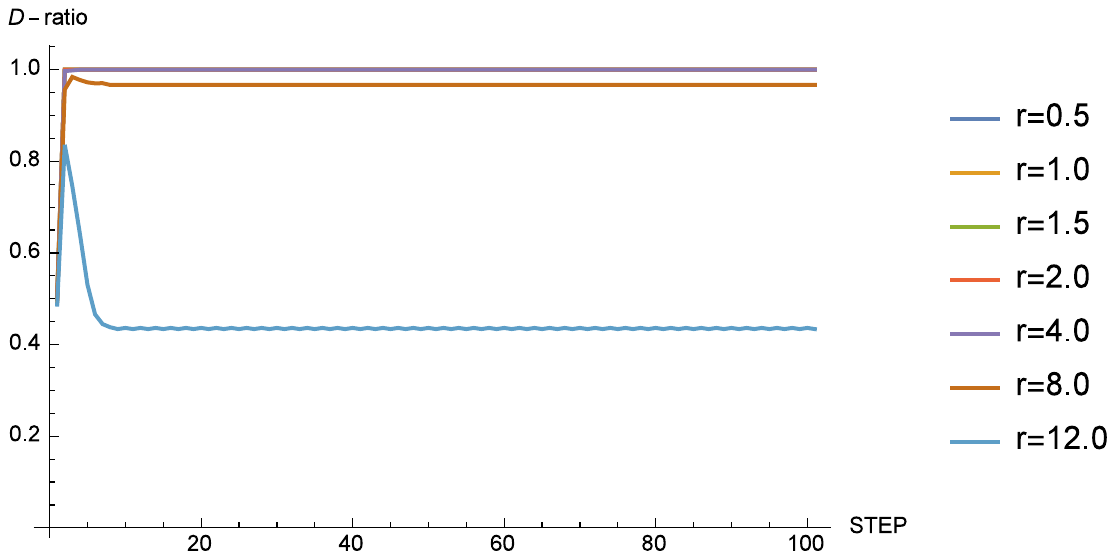}
 \end{center}
 \caption{\small The D ratio in B1-model (WSnet 
 $w=0.01$ and $k=16$) without topology change.}
\label{fig:four}
 \end{minipage}
 \hspace*{3mm}
\begin{minipage}{0.5\hsize}
\begin{center}
\includegraphics[width = 7.0cm,height=3.5cm,clip]{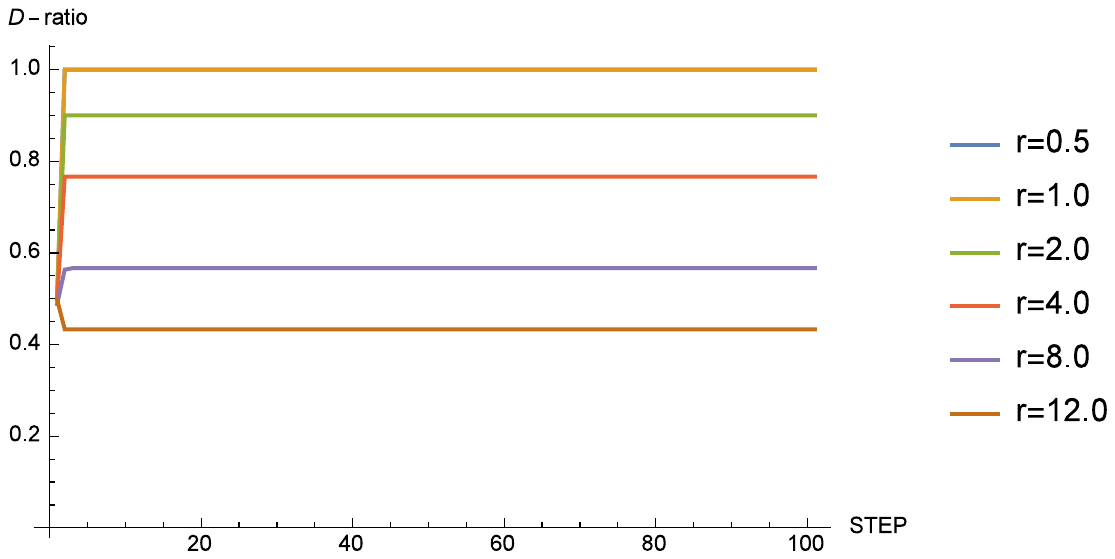}
\end{center}
\caption{\small The D ratio in B1-model (SFnet 
 $k=16$) without topology change.}
\label{fig:two}
\end{minipage}
\end{figure}

 \begin{figure}[btp]
 \begin{minipage}{0.5\hsize}
  \begin{center}
\includegraphics[width = 7.0cm,height=3.0cm,clip]{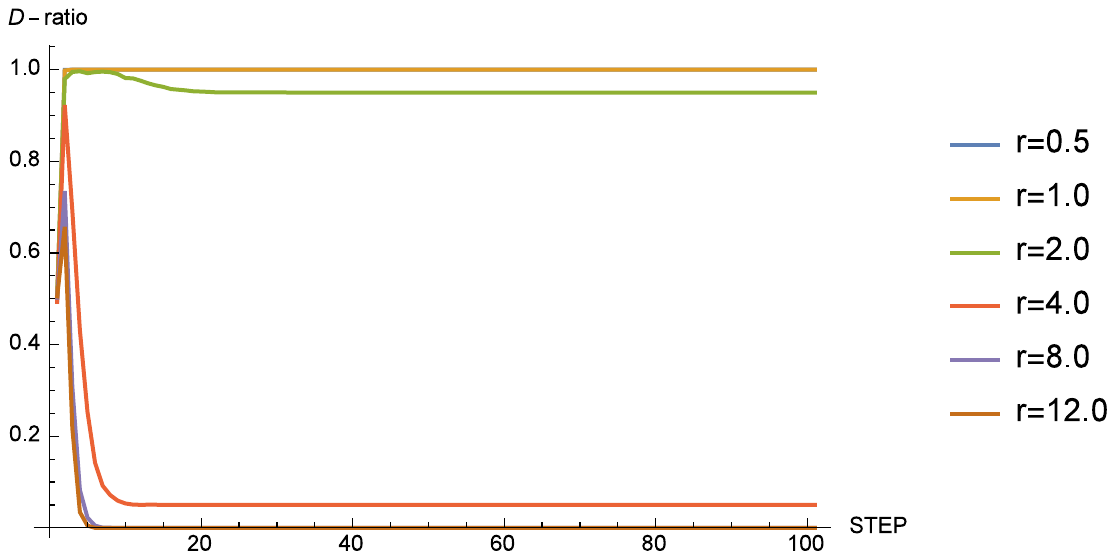}
 \end{center}
 \caption{\small The D ratio in B1-model (WSnet 
 $w=0.01$ and $k=4$) with topology change.}
\label{fig:four}
 \end{minipage}
 \hspace*{3mm}
\begin{minipage}{0.5\hsize}
\begin{center}
\includegraphics[width = 7.0cm,height=3.0cm,clip]{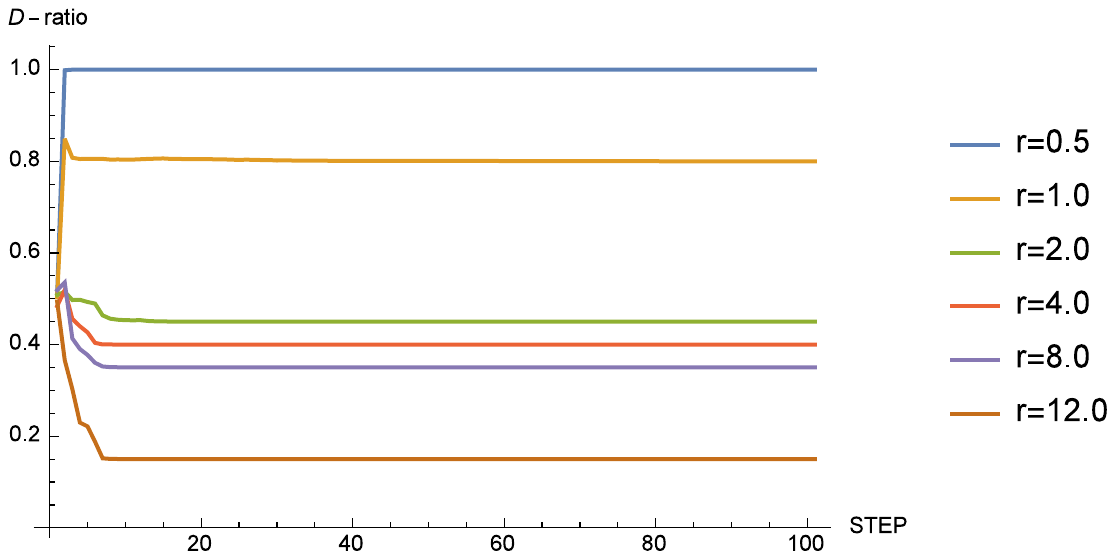}
\end{center}
\caption{\small The D ratio in B1-model (SFnet 
 $k=6$) with topology change.}
\label{fig:two}
\end{minipage}
\end{figure}

% \begin{figure}[tbp]
% \begin{minipage}{0.5\hsize}
%  \begin{center}
%\includegraphics[width = 6.0cm,height=6.7cm,clip,]{B3WSERSFcdnotT20.pdf}
  %\end{center}
  % \caption{\small The D-ratio in B3-model for WSnet($k=4$, $w=0.1$, ERnet $k=16$ and SFnet($k=6$)  without topology change}
  %\label{fig:one}
 %\end{minipage}
%\end{figure}

 \begin{figure}[btp]
 \begin{minipage}{0.5\hsize}
  \begin{center}
\includegraphics[width = 7.0cm,height=3.0cm,clip]{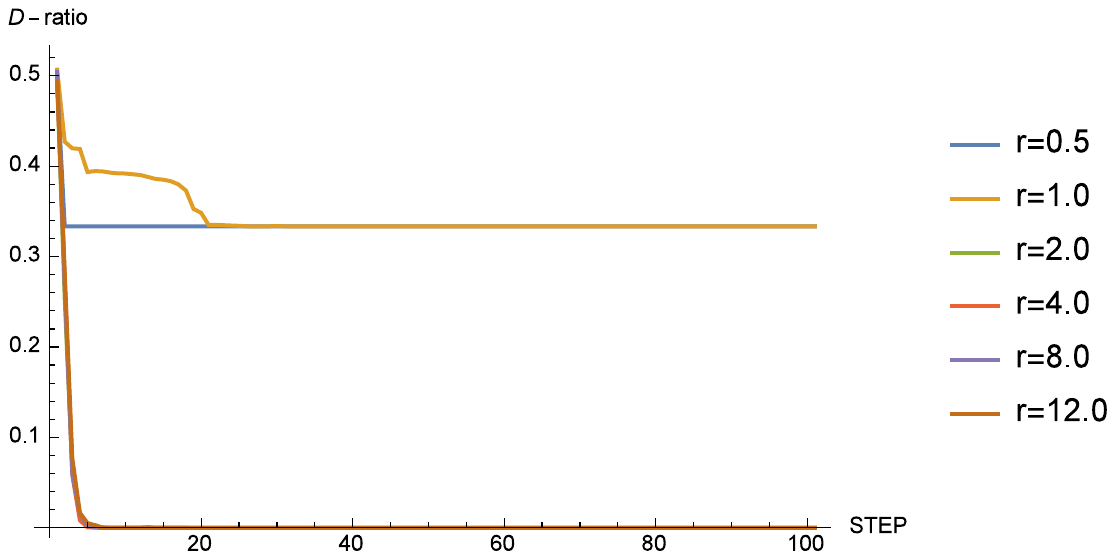}
 \end{center}
 \caption{\small The D ratio in B1-model (ERnet 
 $k=4$) with topology change.}
\label{fig:four}
 \end{minipage}
 \hspace*{3mm}
\begin{minipage}{0.5\hsize}
\begin{center}
\includegraphics[width = 7.0cm,height=3.0cm,clip]{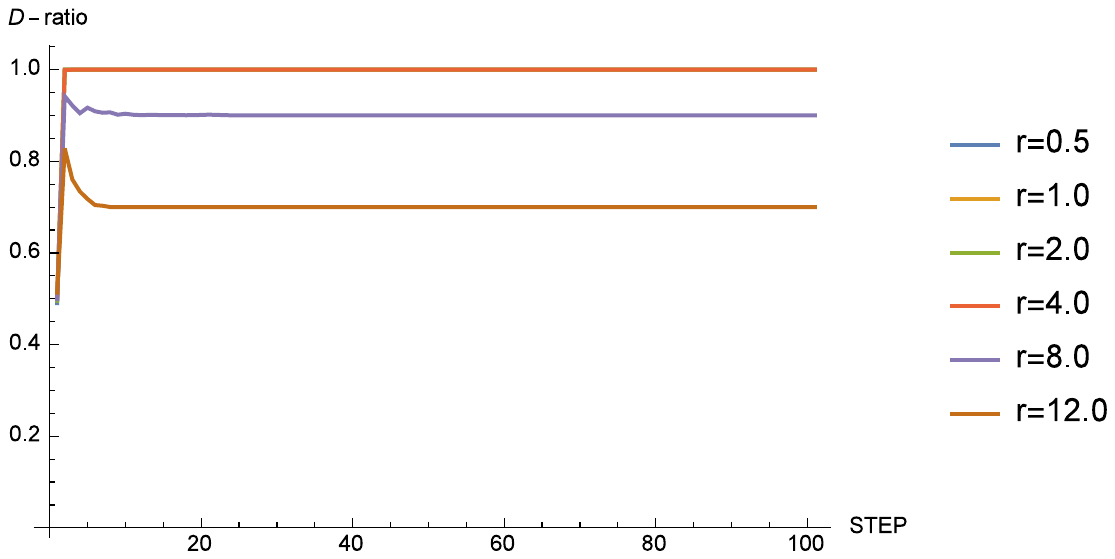}
\end{center}
\caption{\small The D ratio in B1-model (WSnet 
$w=0.01$ and  $k=16$) with topology change.}
\label{fig:two}
\end{minipage}
\end{figure}

\begin{figure}[btp]
\begin{minipage}{0.5\hsize}
\begin{center}
\includegraphics[width = 7.0cm,height=3.0cm,clip]{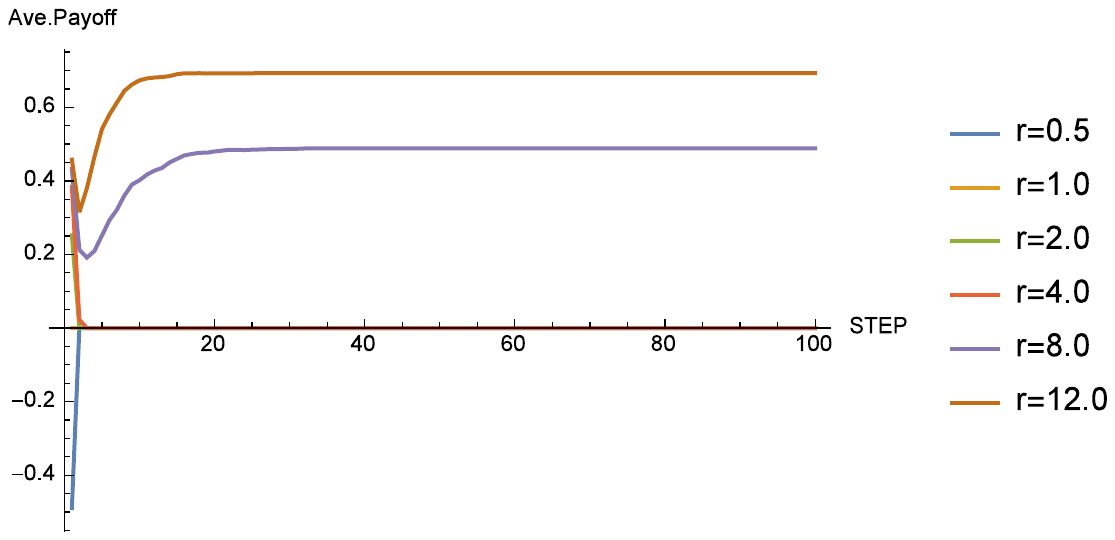}
\end{center}
\caption{\small The average payoff in B1-model (ERnet 
 $k=16$) with topology change.}
\label{fig:two}
\end{minipage}
 \hspace*{3mm}
 \begin{minipage}{0.5\hsize}
  \begin{center}
\includegraphics[width = 7.0cm,height=3.0cm,clip]{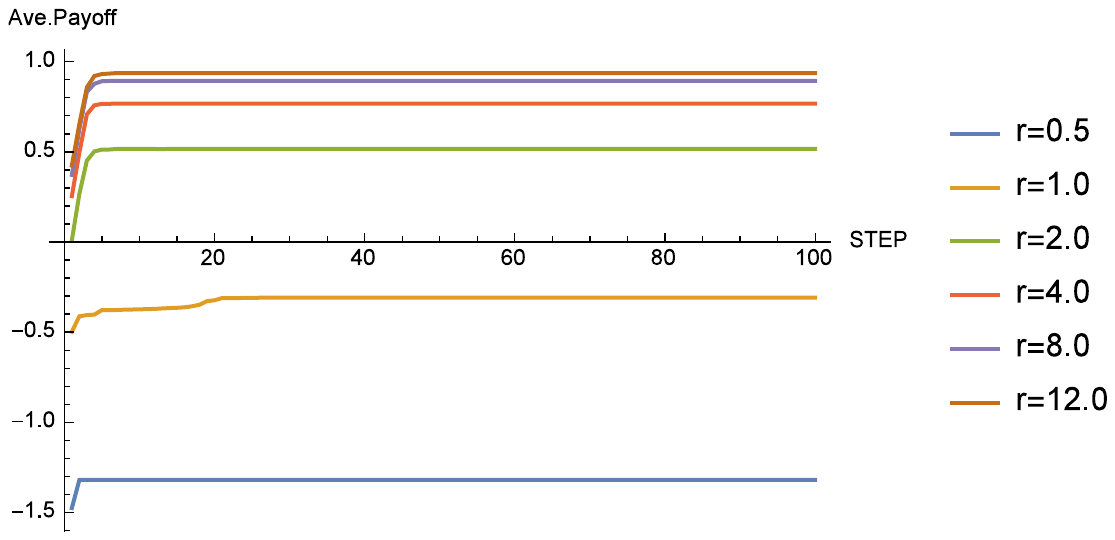}
 \end{center}
 \caption{\small The average payoff in B1-model (ERnet 
 $k=4$) with topology change.}
\label{fig:four}
 \end{minipage}
\end{figure}
 %\begin{figure}[tbp]
 %\begin{minipage}{0.5\hsize}
 % \begin{center}
%\includegraphics[width = 6.0cm,height=6.7cm,clip,]{B3ERpaySLkTnotT23.pdf}
 % \end{center}
  % \caption{\small The average payoff in B3-model for ERnets with topology change ($k=4$) and   without topology change($k=16$) }
  %\label{fig:one}
 %\end{minipage}
%\end{figure}

%%%%%%%%%%%%%%%%%%%%%%%%%%%%%%%%%%%%%%%%%%%
%%%%%%%%%%%%%%%%%%%%%%%%%%%

%%%%%%%%%%%%%%%%%%%%%%%%%%%
\textbf{B-2model:}
%%%%%%%%%%%%%%%%%%%%%%%%%%

The results are essentially the same as those of the previous models\cite{Toyo2}, since this model  does not depend on the game dynamics, 
 but the studies are rather refined in this article. 
 Though D-strategy is superior to C-strategy in  many cases without topology change, regardless of topology change or not which are shown in Fig.30 and Fig.31,    (f30 f31)
 partly there are the cases that C-strategy promotes. 
 A example is shown in Fig. 32(32) whose behavior is like the previous models\cite{Toyo2}. 
 This result presents a  contrast to A2-model independent of game dynamics where topology change has an influence on the promotion of C-strategy. 
 The manifest effect of topology change is not observed in these models adopting fixed strategy. 
 So it is considered that some interaction between topology change and strategy change rule takes place. 
The behaviors of the average payoff are negative only at $r=0.5$  in the all cases. 
An example of these is given in Fig.33. (f33)

%Unlike the previous model \cite{Toyo2}, this model where cooperators contribute a fixed amount per member of the group yield  definite results. 
%Since this model does not depend on game dynamics, the results is  expected to be the same as ones of the previous studies\cite{Toyo2} where definite results were not given. 
%Now we redefine the simulations to insure some definite results. 
%In all initial network models with topology change D-strategy is dominant over C-strategy as shown in Fig.15 and Fig.16, especially at small $k$.
%The increment of D strategy, however, have also a tendency to become mild when $k$ is large in this model.
%As for the network models without topology change, C-strategy sometimes dominates at large $r$ in initial WS networks, which is observed in Fig.17.  
%D-strategy, however, dominates  over C-strategy on the whole in other initial networks as shown in Fig.18.   
%Payoffs are negative in all cases even in the cases of C-dominant as shown in Fig.19. 
%It may be laid down as a general rule that D-strategy, that is to say "defection", prevails and everyone falls  into poverty  in the world where players accommodate themselves to the people about the ones.  
%The  topology change have  somewhat influence on the ration of C-strategy in this model as shown in Fig. 15$-$Fig.18. 
\begin{figure}[tbp]
 \begin{minipage}{0.5\hsize}
  \begin{center}
\includegraphics[width = 7.0cm,height=3.5cm,clip]{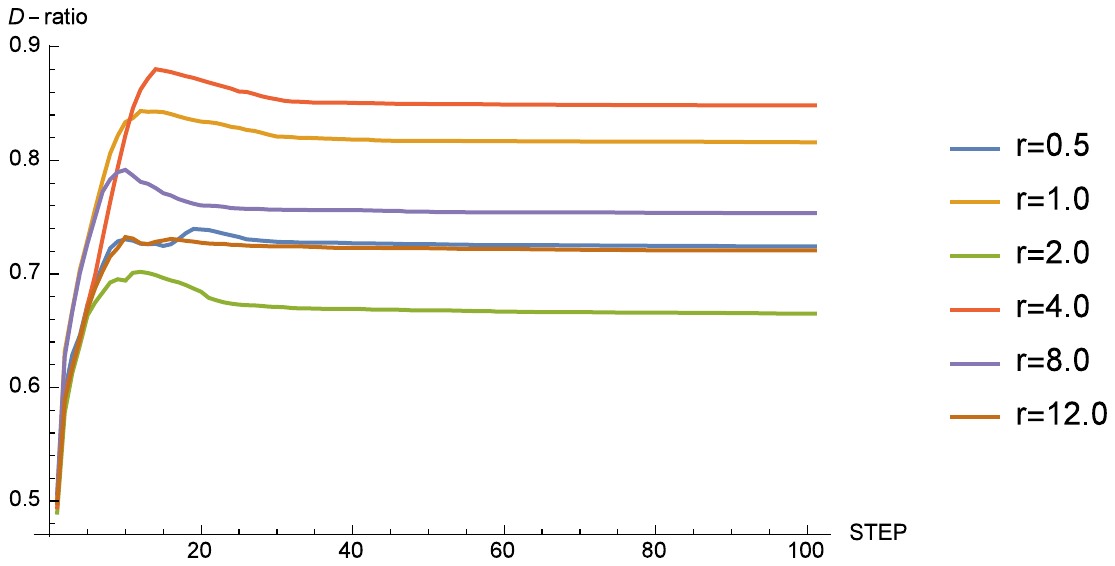}
 \end{center}
 \caption{\small The D ratio in B2-model (WSnet $w=0.1$ and $k=4$) with topology change.}
\label{fig:four}
 \end{minipage}
 \hspace*{3mm}
\begin{minipage}{0.5\hsize}
  \begin{center}
\includegraphics[width = 7.0cm,height=3.5cm,clip]{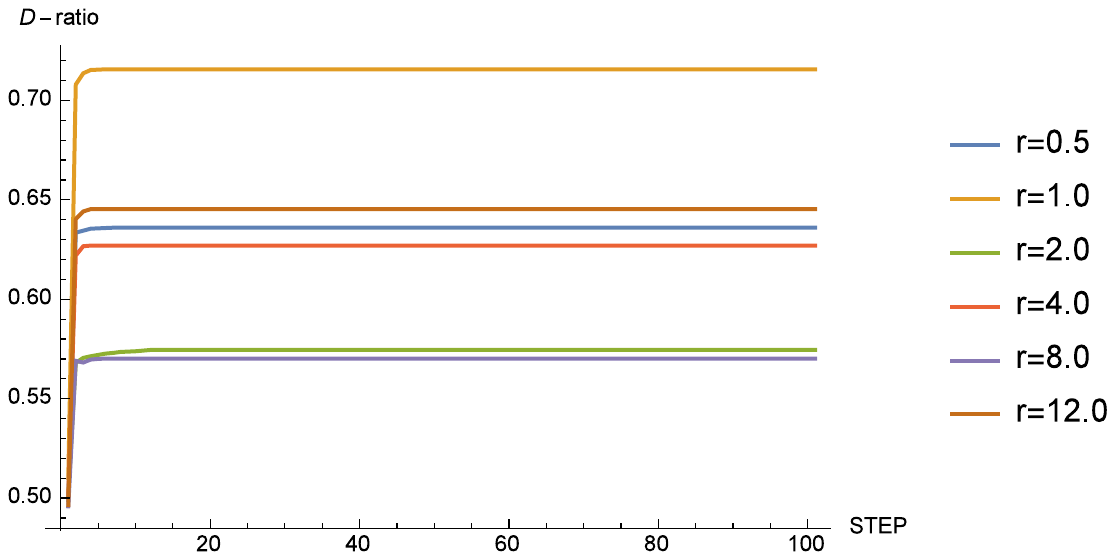}
 \end{center}
 \caption{\small The D ratio in B2-model (WSnet $w=0.1$ and $k=4$) without topology change.}
\label{fig:two}
\end{minipage}
\end{figure}

\begin{figure}[tbp]
 \begin{minipage}{0.5\hsize}
\begin{center}
\includegraphics[width = 7.0cm,height=3.5cm,clip]{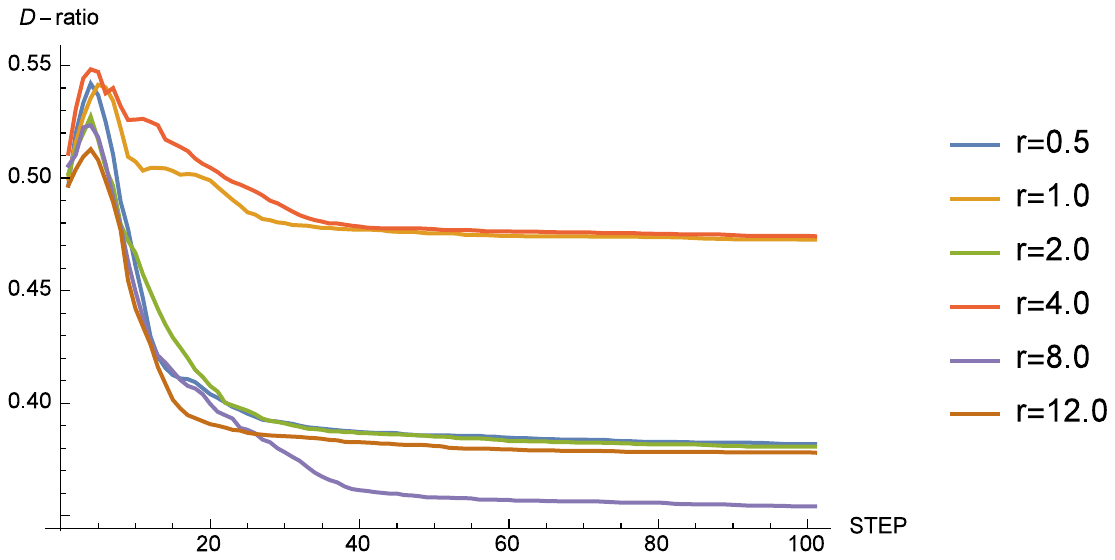}
\end{center}
\caption{\small The D ratio in B2-model (WSnet $w=0.01$ and  $k=4$) with topology change.}
\label{fig:four}
 \end{minipage}
 \hspace*{3mm}
\begin{minipage}{0.5\hsize}
\begin{center}
\includegraphics[width = 7.0cm,height=3.5cm,clip]{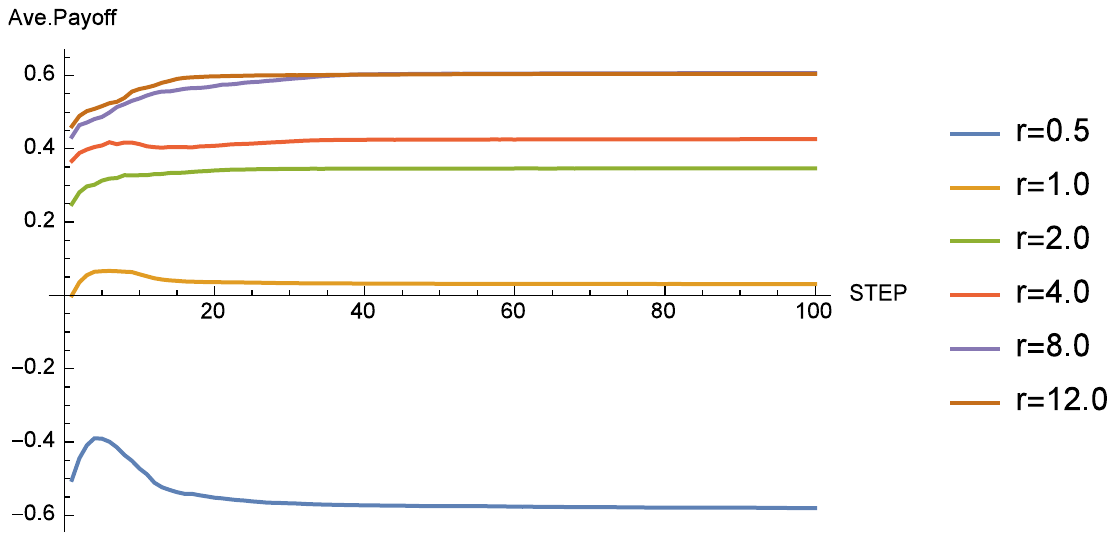}
\end{center}
\caption{\small The average payoff in B2-model (WSnet $w=0.01$ and  $k=4$) with topology change.}
\label{fig:two}
\end{minipage}
\end{figure}

\begin{figure}[!btp]
 \begin{minipage}{0.5\hsize}
  \begin{center}
\includegraphics[width = 7.0cm,height=3.5cm,clip]{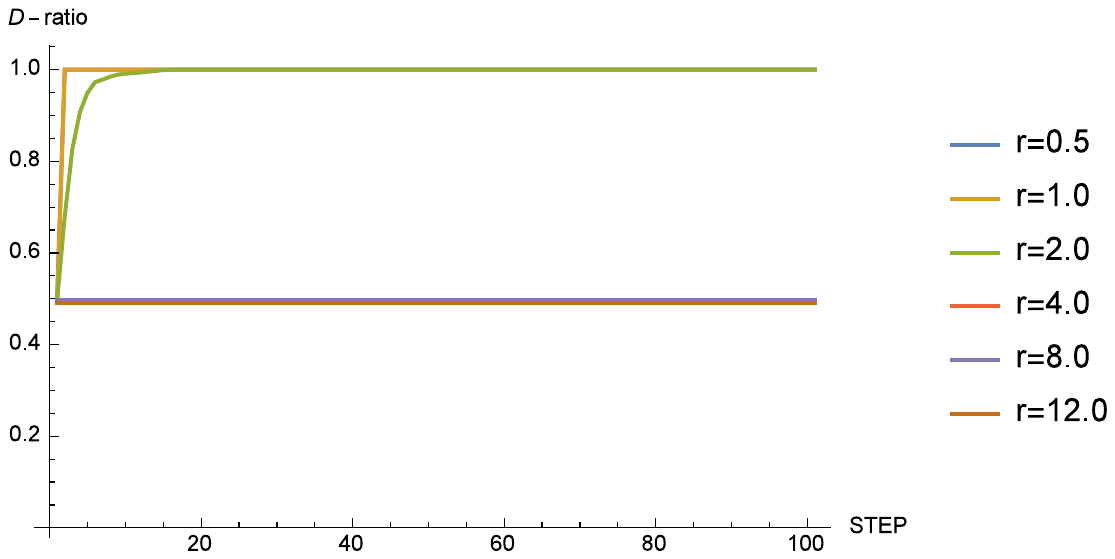}
 \end{center}
 \caption{\small The D ratio in B3-model (WSnet $w=0.1$ and $k=16$) without topology change.}
\label{fig:four}
 \end{minipage}
 \hspace*{3mm}
\begin{minipage}{0.5\hsize}
  \begin{center}
\includegraphics[width = 7.0cm,height=3.5cm,clip]{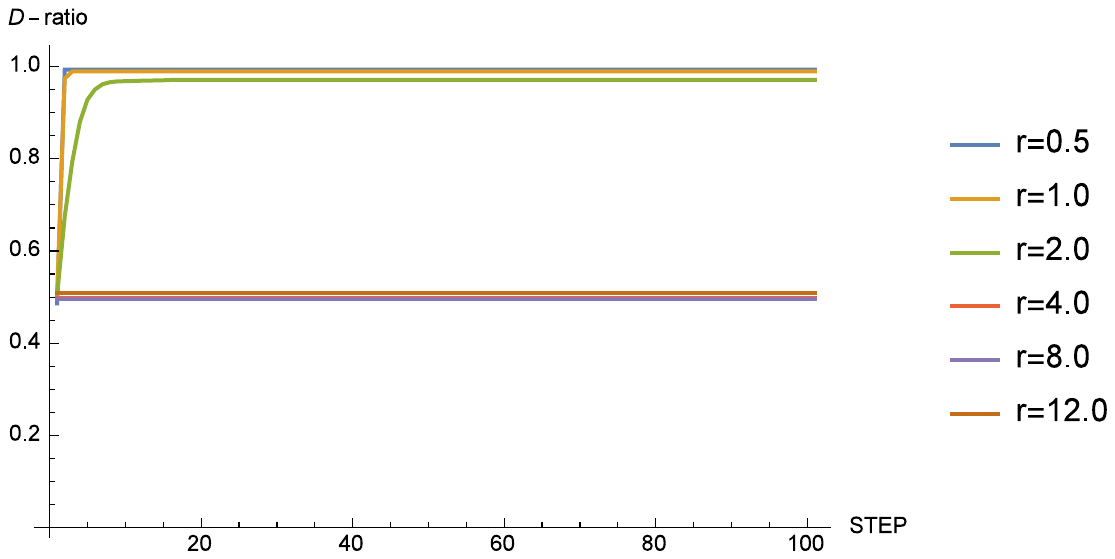}
 \end{center}
 \caption{\small The D ratio in B3-model (ERnet $k=14$) without topology change.}
\label{fig:two}
\end{minipage}
\end{figure}

\begin{figure}[!btp]
 \begin{minipage}{0.5\hsize}
\begin{center}
\includegraphics[width = 7.0cm,height=3.5cm,clip]{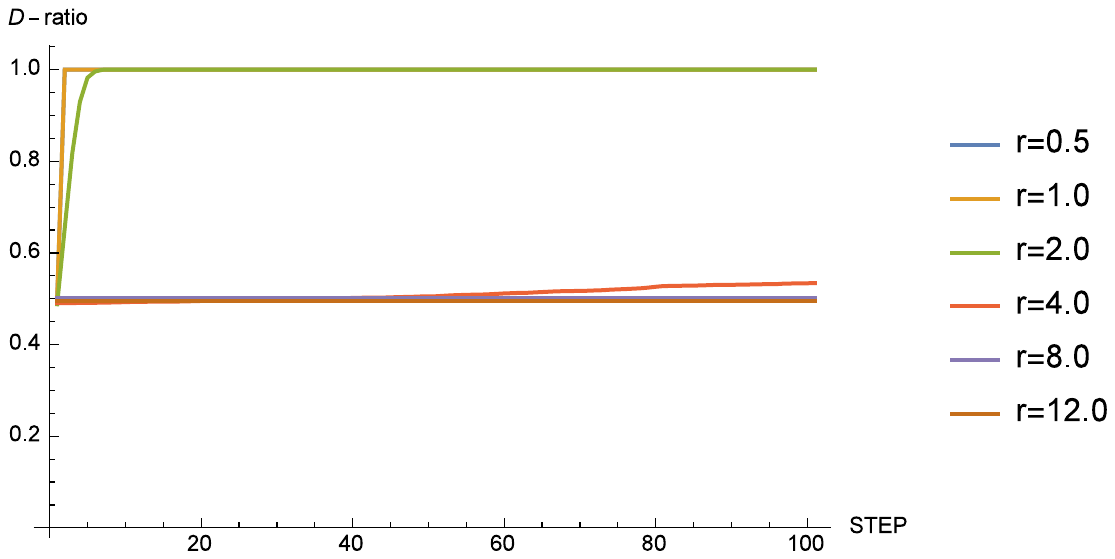}
\end{center}
\caption{\small The D ratio in B3-model (WSnet $w=0.1$ and  $k=16$) with topology change.}
\label{fig:four}
 \end{minipage}
 \hspace*{3mm}
\begin{minipage}{0.5\hsize}
\begin{center}
\includegraphics[width = 7.0cm,height=3.5cm,clip]{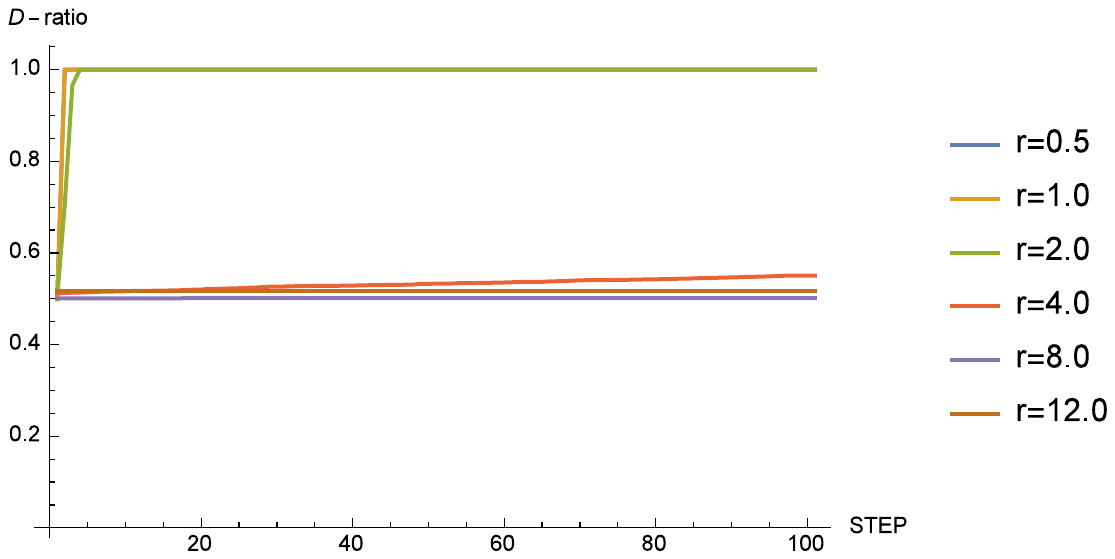}
\end{center}
\caption{\small The average payoff in B3-model (ERSnet $w=0.1$ and  $k=16$) with topology change.}
\label{fig:two}
\end{minipage}
\end{figure}

%%%%%%%%%%%%%%%%%%%%%
\textbf{B-3model:} 
%%%%%%%%%%%%%%%%%%%%%%%
The rate of D-strategy is $100$ percent at small $r$ and about $50$ percent at large $r$ ($r>1$) in about all cases, regardless of whether with topology change or not.  
Thus when the synergy  parameter $r>1$, that is to say, the total token in a pot are multiplied by a factor bigger than $1$, 
the players employing C-strategy and the ones employing D-strategy  coexist at the almost same rate. 
They are a series of figure, Fig.34-Fig.37. ( f34,35,36,37)
Otherwise the players adopting D-strategy sweep across the world.  
This result    presents a  contrast to  the previous model \cite{Toyo2} where C-strategy is promoted when topology change  is not considered.  
But notice that the results in this model are the same as  the previous model \cite{Toyo2}  when topology change is considered. 
The average payoff is positive in almost all cases. 
This is almost the same as other models and thus it is thought that it does not depend on game dynamics and topology change. 

%At the only middle value of $r$, for examples $r=2$ and $r=4$, in the cases witout toology change,
% C-strategy is promoted. 
%The ratio of C-strategy and D-strategy are almost the same in  other values of $r$. 
%They are partly presented in Fig.22.   
% When $k$ is large, almost same results are also given  in the cases with topology change as shown in Fig.21.  
%When $k$ is small, on the contrary, C-strategy is promoted at large $r$ ($r>1$)  in the cases with topology change as in shown Fig.22.    
%Thus the  topology change as well as the average degree have also an influence on the ration of C-strategy. 
%These results turned out the opposite of the results of the model that a player i contribute to a fixed amount \cite{Toyo2}. 
%In general the C-strategi were rather promoted in this model.
%The average payoff is negative for D-dominant and positive for C-dominant in general as shown in Fig.23. 

 %%C-strategy is more promoted only in the cases which topology change does not occur.  
%%Though D-strategy has a majority in the model with topology change  for all $r$ as seen in Fig.16, 
%%C-strategy gets a majority for some $r$ in the cases without topology change in Fig.17. 
%%This behavior is contrary to A-models. 
%%We can not, however, observe the essential differences in both models for the average payoffs as seen in Fig.18 and Fig.19.

%%%%%%%%%%%%%%%%%%%%%%%%%%%
\subsection{Degree Distribution}
%%%%%%%%%%%%%%%%%%%%%%%%%%%%
In models with topology change, the final degree distributions are mainly capable of being classified into four types, as follow. 
They are P, Q, R and pP and their typical examples are given in Fig.38-Fig.41 where P type  represents Poisson like distribution as shown in Fig.38 where dotted data show Poisson distribution.   
pP means the pseud P,  looks like P type but a little different from P type. 
A typical example of pP type degree distribution is given in Fig.39.  
 Q type distribution where many isolated nodes are observed at $r>1$ is given in Fig. 40.  
R type distribution has a sharp peak in the degree distribution as shown in Fig.41. 
An example of  the network structure with Q type degree distribution is given in Fig.42 where overlapped nodes on a straight line in bottom part are all isolated.  
By more refining the simulations than \cite{Toyo2}, we could yield definite results in analysis of this time. 

%%%%%%%%%%%%%%%%%%%%%%%%%%%%%%%%%%%%%%%%%%%%%%%
%%%%%%%%%%%%%%%%%%%%%%%%%%%%%%%%%%%%%%%%%%%%%%%%%%%%%%%%%

In most of A-1 model, the final degree distribution becomes the P or Q type. 
The final degree distribution becomes P type when the average degree $k$ is large and it becomes Q type at small $k$, except for SF network with smal $k$ whose final degree distribution is pP type.
Edges of many nodes with D-strategy is cut from other players in Q type. 
At small $k$,  some nodes  are often isolated by being broken as wrong nodes. 
Once a node is isolated, the node remains to be isolated until the  node is incidentally connected by another node. 
About 20 percent of nodes in many cases remains to be isolated in these cases.  
We can observe it in Fig.40 and Fig.42.  
 However, when the average degree is not so small,  since new edges are  drawn from nodes randomly chosen, it is natural that the final degree distribution comes close to Poisson like distribution.  
%Though only SF net with small $k$ becomes pP..  

In A-2 model, the final degree distribution becomes the pP or Q type. 
when the average degree is large, they are all pP,  and  they are Q at small $k$, except for SF net with small $k$ whose final distribution is pP. 
Then in SF nets, the final degree distribution becomes all pP type, regardless of the average degree in A-2 model. 

In B-1 model, the final degree distributions are all pP type in SF nets and they are all P type in ER net. 
They are P type at large $k$ and pP type at small $k$ in WS nets.  
Thus the final distributions are all Poison or pseudo Poisson distributions in B-1 model. 

 B-2 model show most various properties in the final degree distribution.  
The final degree distributions become P type except for SF net  and WS net with  large  $k$. 
%All the final degree distribution for SF net with small $k$ are pP type. 
The final degree distribution in SF net with large $k$ shows an odd property that is a mixture of Q and R.   
So we call it QR type. 
They are all P type in ER nets except A-2 model at large $k$.   
In WS nets,  though the final degree distribution is P at small $k$,  it is R type at large $k$ regardless of $w$, and R type appears only in this case.  
R type distribution means that the frequency of rewiring is small or nodes with nearly same number of edges mainly rewire together.      
Thus the  peculiar types of final distributions appear in B-2 model. 

In B-3 model,  the final degree distributions become Poisson like distribution  in the all cases. 
In B3 model it is easy that the flip  from C-strategy to D-strategy happens and the edge shot out from the node(player) of the D-strategy is often broken from other nodes. 
So it is considered that the random connection to new nodes leads to Poisson like degree distribution when rewiring often happen in B3 model.  

Though the degree distributions of initial SF nets, especially at small $k$,  are almost pP type (really an exception happens only in B-3 model),    
it is almost P or pP at large $k$.  
Thus the final degree distributions have mostly settled into  Poisson like distribution, that is to say P and pP.  
Q type mainly appears in WS nets and ER nets with small $k$ 
 
 As a whole, P type and pP type distribution are observed in comparatively large number of times which means that the rewiring rule  in the present model, where a new node at rewiring  is chosen from random nodes, has a great  influence on the final degree distribution. 
These results stand in contrast to the models that a player contribute to a fixed amount, where the final degree distribution were Poisson like in almost all cases except for Q type which sometimes happend\cite{Toyo2}. 
%%%%%%%%%%%%%%%%%%%%%%%%%%%%%%%%%%%%%%%%%%%%%%%
%%%%%%%%%%%%%%%%%%%%%%%%%%%%%%%%%%%%%%%%%%%%%%%%%%%%%%%%%

%Meanwhile, the final degree distribution is P type in most of cases of B2 and B3 except for B2 model in the WS %networks with large $k$.  
%The initial degree distribution plays crucial role in the final degree distribution of the B1 model but is independent of the average degrees. 
%The final degree distribution is P type for ER networks,  R type for WS networks and pQ type for SF networks in B1 model.      
%R type arises for B1 model in WS network and also for B2 model with large $k$ in WS network. 
%Then R type means that the drgree disteibution does not remarkably change from initial degeree distribution of WS network.  
%The degree distributions in B3 model are P or pP type. 

%%%%%%%%%%%%%%%%%%%%%%%%%%%%%%%%%%%%%%%%%%%%%%%%%%%%%%%%%%%%%%%%%%%%%%%%%%%%%%%%%%%%%%%

\begin{figure}[btp]
 \begin{minipage}{0.5\hsize}
\begin{center}
\includegraphics[width = 7.0cm,height=3.5cm,clip]{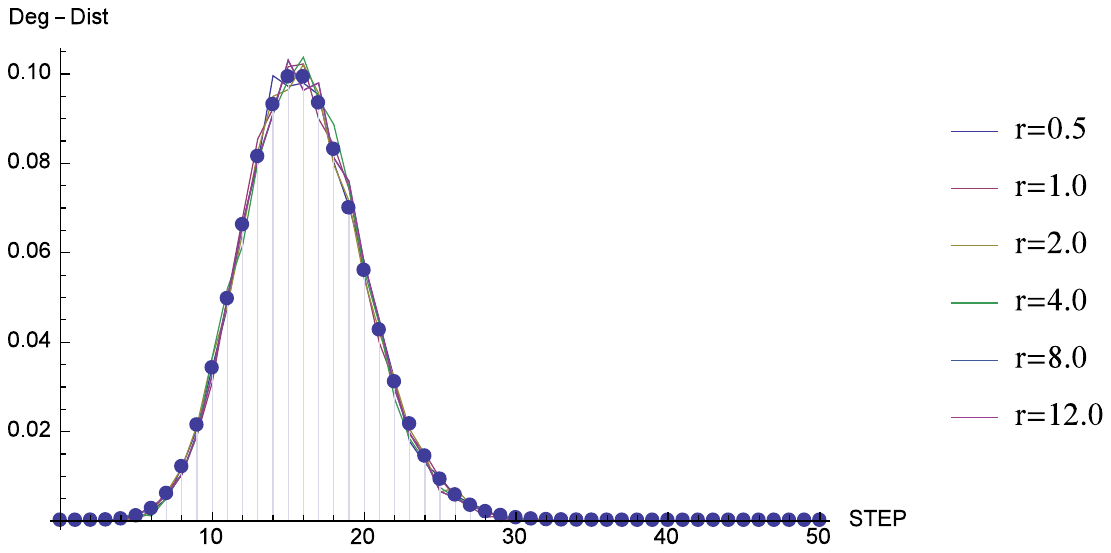}
\end{center}
\caption{\small The final degree distribution in A1-model (WSnet $w=0.01$ and  $k=16$) with topology change.}
\label{fig:four}
 \end{minipage}
 \hspace*{3mm}
\begin{minipage}{0.5\hsize}
\begin{center}
\includegraphics[width = 7.0cm,height=3.5cm,clip]{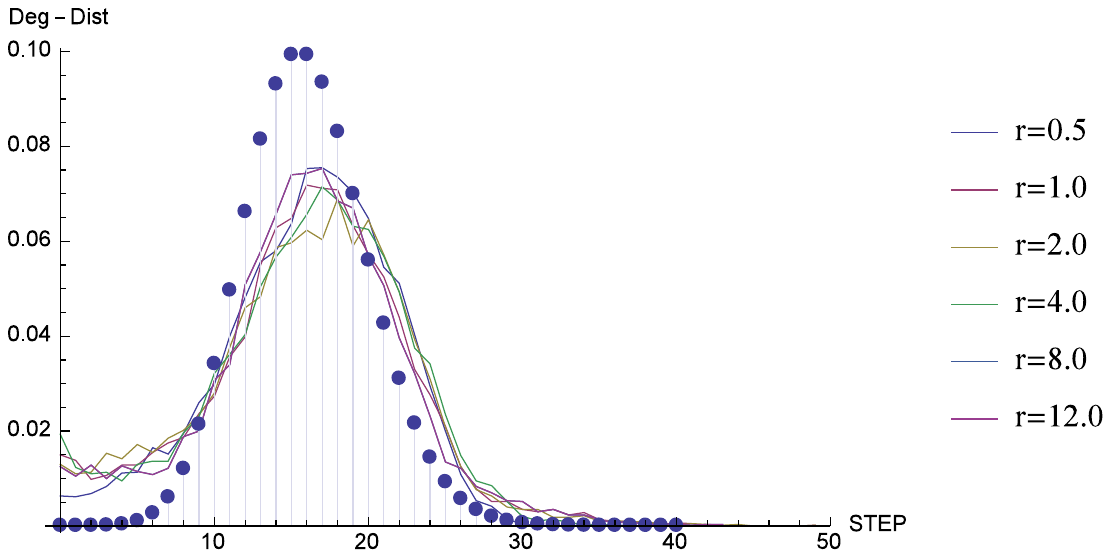}
\end{center}
\caption{\small The final degree distribution in A3-model (WSnet $w=0.0$ and  $k=16$)  with topology change.}
\label{fig:two}
\end{minipage}
\end{figure}

\begin{figure}[btp]
 \begin{minipage}{0.5\hsize}
\begin{center}
\includegraphics[width = 7.0cm,height=3.5cm,clip]{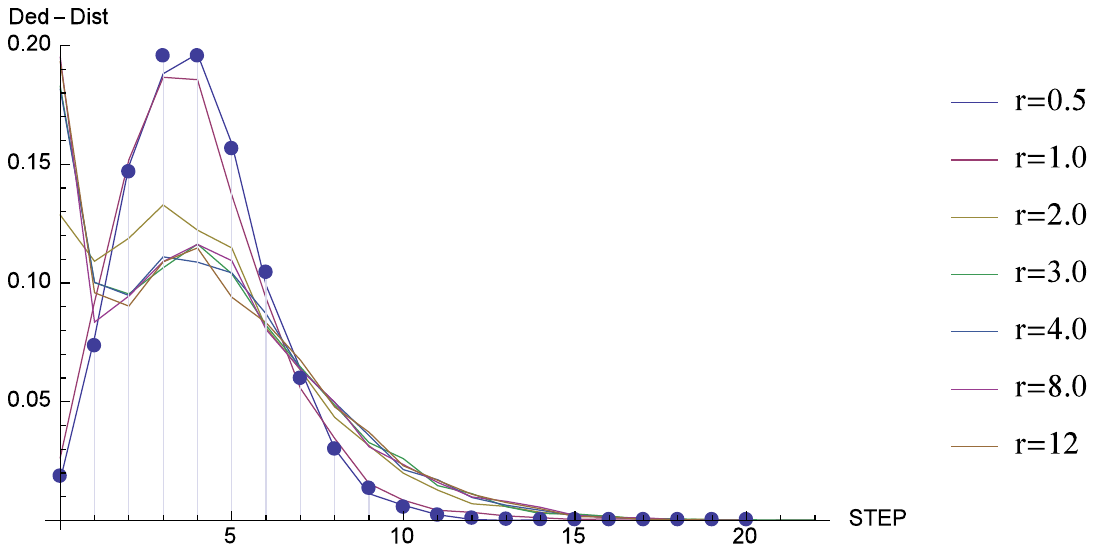}
\end{center}
\caption{\small The final degree distribution in A1-model (WSnet $w=0.1$ and  $k=4$) with topology change.}
\label{fig:four}
 \end{minipage}
 \hspace*{3mm}
\begin{minipage}{0.5\hsize}
\begin{center}
\includegraphics[width = 7.0cm,height=3.5cm,clip]{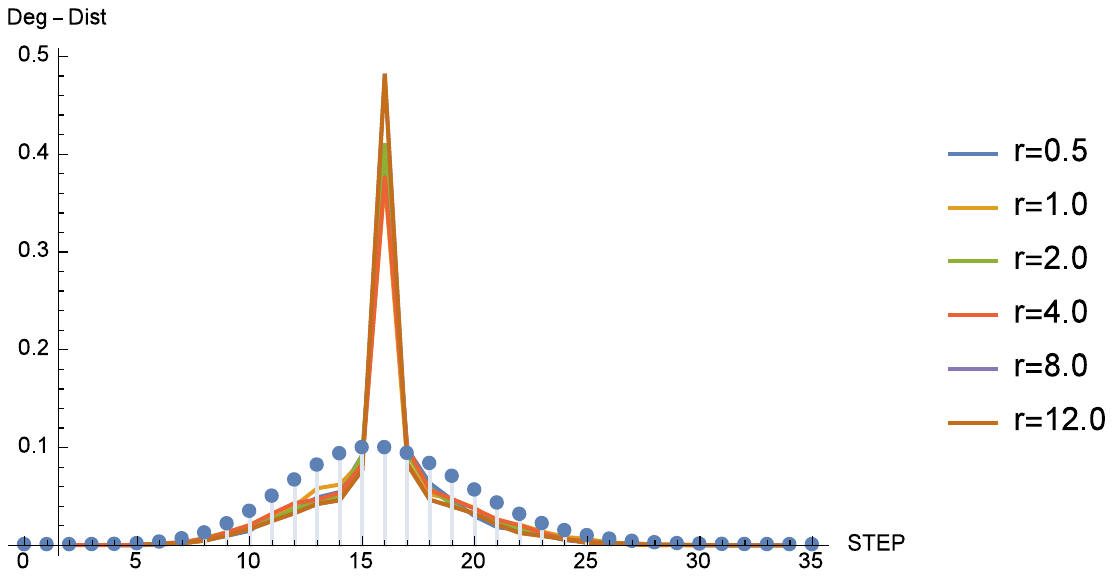}
\end{center}
\caption{\small The final degree distribution in B2-model (WSnet $w=0.01$ and  $k=16$)  with topology change.}
\label{fig:two}
\end{minipage}
\end{figure}

\begin{figure}[btp]
 %\begin{minipage}{0.5\hsize}
\begin{center}
\includegraphics[width = 5.0cm,height=4.5cm,clip]{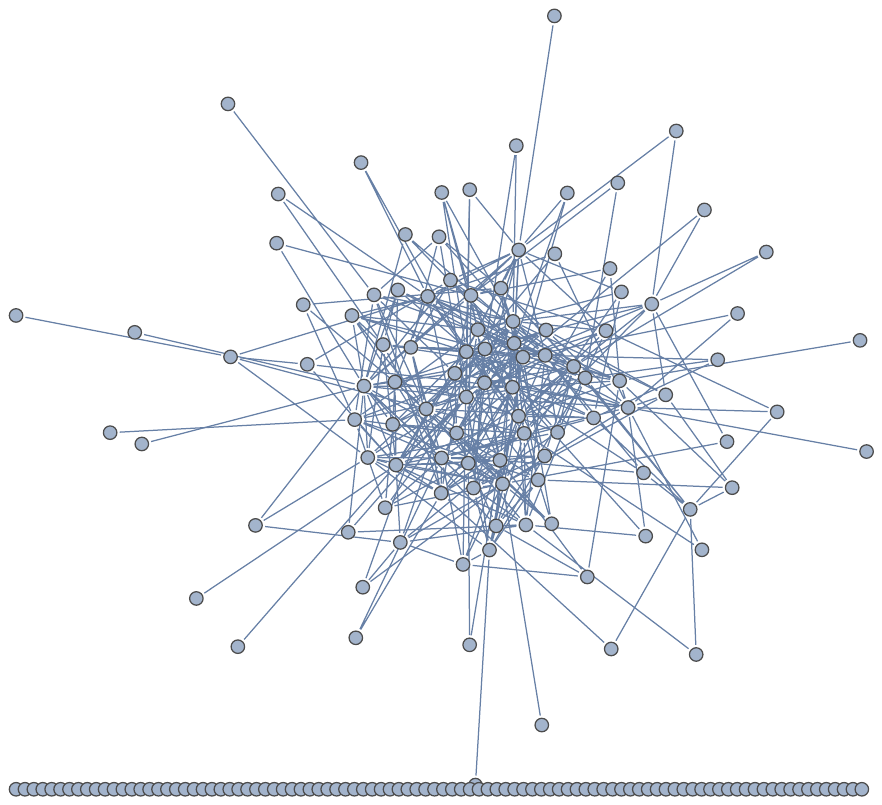}
\end{center}
\caption{\small The network structurein A1-model (ER net  $k=4$) with topology change.}
\label{fig:four}
 %\end{minipage}
\end{figure}

%Since the average payoff and D-ratio in $\beta$-models were almost the same as in $\alpha$-models, we omit the details of them. 
%But there are apparent differences in the degree distributions in the both models. 
%These are shown in Fig.22 and Fig.23. 
%The degree distributions at small $k$ basically become line graphs shown in Fig.22 in A-models and also B1-model 
%where players with D-strategy who account for a large portion are almost  isolated and have zero degree. 
%On the other hand, at large $k$ many players with D-strategy are often isolated but the degree distribution has a small peak at the  original average degree in B2-model and also B3-model. 
%Such the situations are not the characteristic of WS-net.  
%These characteristics do not depend on initial networks. 
%In $\beta$-model, when a player does not have the neighbor of his/her neighbor, the total number of edges decreases. 
%Especilally it is easy to fall into such situation for players with D-strategy, who are often broken with other players. 

%%%%%%%%%%%%%%%%%%%%%%%%%%%%%%%%%%%%%%%%%%%
%%%%%%%%%%%%%%%%%%%%%%%%%%%%%%%%%%%%%%%%%%%
%game dy> top change, small k large r
%%%%%%%%%%%%%%%%%%%%%%%%
\subsection{Promotion of C-strategy and Enhancement of Payoff}
%%%%%%%%%%%%%%%%%%%%%%%%%%

Though undermentioned all findings in this section were not necessarily shown by the figures in this article, 
we summarize the cases in which C strategy is promoted. 

(1)The most of A1 models, regardless of $r$,  with topology change, except for SF networks. \\

(2)The most of A2 models at large $r$, regardless of considering topology change or not.

 \hspace{5mm}Especially,  C-strategy is remarkably promoted at all $r$ in A-2 model with topology change.\\

(3)B1 models with small $k$ at large $r$, regardless of considering topology change or not.

\hspace{5mm}Contrary to small $k$,  D-strategy is promoted at large $k$.  \\

Notice that C-strategy is not promoted in the almost cases of B-2 and B-3 models. \\ 
The topology change affects the promotion of C strategy in A models to some degree, especially the model with topology change definitely promotes C strategy in A2 model.  
Thus the case that a player plays with different strategies with individual players is rather promoted than with the same strategy in general. 
Especially the C-strategy is  easier to be promoted in the case that a player  strategically accommodates to his/her friends which  link with the player  than the case that the player imitates the strategy of the player that gets the largest payoff in the players connecting to the player.

There is not crucial effects of topology change on promoting C-strategy in B models. 
Contrary to A-2 model, C-strategy is not promoted B-2 model where players are opportunism. 
Moreover  C-strategy is not also promoted in the models where players avoid a loss. 

%Meanwhile, there is some effect on promoting C strategy in B3 model, but  the average degree in the initial networks has more  influence on promoting C strategy than topology change. 
On the whole, in the cases with large $r$ and small $k$ C-strategy  is more promoted than when it is not so. 
It is natural that C-strategy is promoted at large $r$, because the chance that a player adopting C-strategy  suffers  a loss is low. 
 Players that have a little friends have a tendency to adopt C-strategy gradually. 
The reason is considered that it is of much benefit when the player that has a little friend breaks the relation with a player adopting D-strategy. 

While C-strategy is not necessarily promoted, there are some cases where C-strategy   coexist with D-strategy in nearly equal proportions. 
They are\\

(I) A-1 model without topology change and A-1 model in SF net, \\
   
(II)A-2 model at large $k$  for some $r$,\\

(III)B-1 model  without topology change and with topology change for some $r$,\\

(IV)B-3 model for large $r$,\\

These cases are significant in the meaning that D-strategy, which is rational from game theoretical point of view, does not sweep across all players.

The rough summarization in the relation between the promotion of C-strategy and topology change and game dynamics is as follows.
C-strategy is also promoted in some cases in the models which reflect game dynamics, particularly the cases at small $k$ and large $r$. 
 C-strategy is more promoted in A1 model with topology change than that without topology change. 
Thus the  topology change is effective somewhat in promoting  C-strategy but its effect is without much effect.  
Game dynamics, however, has more effect than topology change on promoting C strategy in general, as shown the results in A-1 model and B-1 model.
Moreover we could find the synergistic effect in promoting C-strategy in A1 model to some degree.
Then taking things by and large, small $k$ has more potential to promote C-strategy than  large $k$.% when it is not so. 

  As for payoff,  it  takes the value between $0.0\sim0.5$ or  $0.0\sim1.0$ for $r\geq 1.0$ but about $-0.4$ for $r<1$ in B models. 
  These results do not depend on $k$, initial networks and whether topology change is considered or not. 
The more $r$ is large, the more the average payoff is also large.   
Of course, it is natural.  
 Most of A models are in similar results. 
However,  the average payoff takes  of $0.0\sim \; tens\;\mbox{or } 100 $  in SF nets in A-1 model. 
It  takes  of $0.0\sim \; tens $  in WS nets in A-2 model.
It is not clear why  such curious behaviors in A models happens as yet.    
As a result, we can not observe  drastic enhancement in payoffs except for a few cases.

%Furthermore the average payoff is negative when the ratio of D strategy is  not less than $0.5$ and 
%positive when that of C-strategy is not less than $0.5$ in all models. 
%So when all cooperate, the whole world would more flourishes than when it is not so.
%AAAAAAAAAAAAAAAAAAAAAAAAAAAAA%%%%%%%%%%%%%%%%%%%%%%%%%%%%%%%%%%%%%%%%%%%

 % \end{document}

%%%%%%%%%%%%%%%%%%%%%%%%%%%%%%%%%%%%%%%%%%%%%%%%%%%%%%
\section{Summary}
  In this article, we consider an coevolutionary PGG on diverse complex networks where the topology of the networks varies mainly under the influence of game dynamics.  
The diverse tactics are considered to model simply the actions that usual persons would take. 
Thus the models presented in this article are so simple to investigate only the  essential skeleton of  properties in PGG on networks.   
Moreover we tried to make models that do not include any factors which explicitly promotes cooperation in order to study the effect of only topology change and game dynamics on cooperation , as far as I can do it. 
Thus we tried to exclude the factors that cooperation is plainly promoted from model building. 
The only potential that may promote cooperation is to break the edges arising from nodes(players) with D-strategy in the presented models.     
The some models reflect game dynamics whereas others  do not in the change of strategies and the network structure. 
Diverse complex networks are investigated as initial networks. 

In this research, we studied the ratio of players with D-strategy, the average payoff over all players and the final degree distributions by simulating  coevolutionary PGG, where players have various types of tactics, on complex networks. 
 We investigated whether the interaction between the game dynamics and topology change, or a combined effect of them promotes cooperation and  make people (players) wealthy or not.  

The similar studies have been already investigated in \cite{Toyo2}.  
Though these cooperators can contribute a fixed amount per game in PGG on networks, cooperators contribute a fixed amount per member of the group in PGG of this article. 
This is a great difference in  both PGGs. 
The model in this article is considered to be more realistic than the previous model, since it is assumed implicitly that players have some finite asserts. 

As a result, models with game dynamics or some strategy change rules  and the models with topology change promote C-strategy, respectively. 
However, the game dynamics and the topology change do not produce a marked  synergy effect. 
An exception that the combined effect somewhat promotes C-strategy is only A-1 model  (and B1-model with small $k$ and large $r$.) 
 C-strategy, however, rather tends to comparatively flourish, especially when $k$ is small than when large $k$.  
The reason is that when $k$ is small, the possibility that is  damaged by D-strategy will diminish, because the effect that a player breaks the edge to a player with D-strategy at small $k$ is more showing up.   
As for large $k$,  the effect is rather light. 
 %%%%%%%%%%%%%%%%%%%%%%%%%%%%
%%%%%%%%%%%%%%%%%%%%%%%%%%%%
%%%%%%%%%%%%%%%%%%%%%%%%%%

As for the final degree distribution, it depends on the nature of  initial networks  and sometimes the average degree. 
The final degree distribution of the initial SF-net change drastically into Poisson like (P-like) distribution in most cases. 
The small $k$ except SF nets leads to Q type distribution but large $k$ leads to Poisson like ones in A-models. 
In B-models, the final degree distributions are  mainly Poisson like but sometimes become R or Q type. 
Poisson-like distributions mean that frequent rewiring  happens and Q-like one show that there are many isolated nodes. 
R type means that the original degree distribution does not drastically change in WS-net.

%Especially A2-model remarkably promote C-strategy but it is independent of the game dynamics. 
%Thus the  increment of C-strategy is not due to the interaction or the combined effect but only due to topological evolution.   
%In contrast,  
%D-strategy is obviously dominant in the models without topology change and PGG dynamics. 
%Other factor such as punishment for promoting so much C-strategy may be needed\cite{And},\cite{Perc8}.  

The models considered in this article is so simple and have only some typical properties or a part of them of usual people.  
Diverse people, who has more involved characteristics, coexist in the real world. 
We should make more realistic models, where players can employ diverse tactics simultaneously, A1, A2, B1 $\cdots$ and so on, to reflect such the situation. 

 The results in this article are preliminary ones, where some models with tactics probable in the world are considered in a piecewise manner and the  simulation size is not so large  only to uncover some typical results. 
Deeper considerations with more general modeling and large-scale simulations are therefore indicated in subsequent studies.

\end{document}